\documentclass[11pt]{article}

\usepackage {color}

\usepackage {ucs}
\usepackage{upgreek} 						
\usepackage{mathbbol}
\usepackage [utf8x]{inputenc}
\usepackage {graphicx}
\usepackage {enumitem}						
\usepackage {amsfonts}

\usepackage {multicol}						
\usepackage {datetime} 						
\usepackage [FIGTOPCAP]{subfigure}						
\usepackage{sidecap}					

\usepackage {parskip}						
\usepackage {setspace}						
\usepackage[font=small,textfont=it,width=0.97\textwidth]{caption}		
\usepackage {mathtools}						
\usepackage {array}							
\usepackage{tabularx}
\usepackage{changepage} 			
\usepackage{titlesec}						

\newcounter{sectionexp}
\titleclass{\sectionexp}{straight}[\section]
\titleformat{\sectionexp}[runin]
{\normalfont}{\textbf{Experiment \thesectionexp}}{0.5em}{}
\titlespacing*{\sectionexp}{0pt}{*}{*}
\renewcommand{\thesectionexp}{\arabic{sectionexp}}

\renewcommand {\vec}{\mathbf}

\renewcommand{\a}{\alpha}
\newcommand{\R}{\mathbb R}

\renewcommand{\(}{\left(}
\renewcommand{\)}{\right)}

\graphicspath{{./input/}}

\def\subtext#1{\text{#1}}

\title{Modelling cell-cell collision and adhesion with the Filament Based Lamellipodium Model}
\author{
	N. Sfakianakis\thanks{Institute of Applied Mathematics, Heidelberg University, Im Neuenheimerfeld 205, 69120, Heidelberg, Germany,  \mbox{sfakiana@math.uni-heidelberg.de} 
	}
	\and
	D. Peurichard\thanks{Laboratoire Jacques-Louis Lions, INRIA,  Sorbonne University, Place Jussieu 4, Paris, France, \mbox{diane.a.peurichard@inria.fr} 
	}
	\and
	A. Brunk\thanks{Institute of Mathematics, Johannes Gutenberg University, Staudingerweg 9, 55128, Mainz, Germany, \mbox{abrunk@uni-mainz.de} 
	}
	\and
	C. Schmeiser\thanks{Faculty of Mathematics, University of Vienna, Oskar-Morgenstern-Platz 1, 1090,  Vienna, Austria, \mbox{christian.schmeiser@univie.ac.at} 
	}
	}

\begin{document}
\maketitle
\begin{abstract}
		We extend the live-cell motility Filament Based Lamellipodium Model to incorporate the forces exerted on the lamellipodium of the cells due to cell-cell collision and \textit{cadherin} induced cell-cell adhesion. We take into account the nature of these forces via physical and biological constraints and modelling assumptions. We investigate the effect these new components have in the migration and morphology of the cells through particular experiments. We exhibit moreover the similarities between our simulated cells and HeLa cancer cells. 
\end{abstract}

\section{Introduction.} \label{sec:intro}
\textit{Cell adhesion} is a key process in a wide range of biological phenomena. It usually acts together with \textit{cell migration} and plays a fundamental role in the development of the organism e.g. during the \textit{gastrulation} and the \textit{patterning} phases of a vertebrates' body. They both remain important even after development for the maintenance and repair of the cell and tissue structure. On the other hand, their dysregulation has been associated to a number of \textit{diseases} and \textit{conditions} including \textit{tumour metastasis}. 

\textit{Cell adhesion} is the result of interactions between specialized proteins found at the surface of the cells termed \textit{cell-adhesion molecules} (CAM). The CAMs are divided into four main groups: \textit{integrins}, \textit{immunoglobulins}, \textit{cadherins}, and \textit{selectins}. Out of these, the \textit{integrins} participate primarily in the \textit{cell-extracellular matrix} (ECM) adhesion and play a pivotal role in the migration of the cells, and the \textit{cadherins} (calcium dependent adhesions) are fundamental in cell-cell adhesion and in the formation of cell clusters and tissues.

The \textit{cadherin} proteins are comprised of three domains, an \textit{intracellular}, a \textit{transmembrane}, and an \textit{extracellular} domain. The \textit{intracellular} domain is linked to the \textit{actin filaments} (F-actin). The extracellular domain binds to the corresponding part of \textit{cadherins} of neighbouring cells. This domain is highly binding specific and accordingly classifies the \textit{cadherins} in several types (\textit{E-}, \textit{N-cadherins} etc.). The expression levels of these \textit{cadherin} types lead to preferential adhesion organization of the cells and to the formation of different tissues.

In the current paper, our objective is to model \textit{cadherin} induced cell-cell adhesion and combine it  with a mathematical model of cell migration and cell-ECM adhesion. We focus on a particular type of cell migration in which the lamellipodium of the cell plays a pivotal role. It is termed \textit{actin-based motility} and is employed by fast migrating cells such as \textit{fibroblasts}, \textit{keratocytes}, and \textit{cancer cells}. 

There have been several efforts to model and simulate this type of cell migration in the literature, see e.g. \cite{Stevens2015, Voigt2015, Mogilner2009, Alt2009, DeSimone2011, Preziosi2013, Merkel2012, Madzvamuse2013, Ambrosi2016, Schwarz2010, Schwarz2012}. Here, we use and build on the \textit{Filament Based Lamellipodium Model} (FBLM). This is a two-dimensional, two phase model that describes the lamellipodium at the level of actin-filaments. The FBLM was first derived in \cite{Oelz2008, Oelz2010a} and later extended in \cite{MOSS-model}. When endowed with a particular problem specific \textit{Finite Element Method} (FEM), the resulting FBLM-FEM is able to reproduce biologically realistic, crawling-like lamellipodium driven cell motility \cite{MOSS-numeric, Brunk2016, Sfakianakis2018}.

Although the FBLM describes the dynamics of the actin-filaments and the lamellipodium, the deduced motility is understood as the motility of the cell. This is primarily because of the predominant role of the lamellipodium in the motility of the model-biological cell (i.e. \textit{fish keratocyte}) that we consider, \cite{Small2002}. So, for the rest of this work we will not distinguish between the two, and we will use the term \textit{cell motility} in both cases.

The extensions of the FBLM that we propose in this work account for two phenomena. The exchange of \textit{cadherin} mediated adhesion, and physical collision forces between two neighbouring cells. These forces are attractive in the case of cell-cell adhesion and repulsive in the case of cell-cell collision. They are introduced in the FBLM via an attractive-repulsive potential that depends, non-linearly, on the relative distance of the two cell membranes. When the cells come close enough, within a \textit{cadherin} length distance, an attractive force is developed between the two membranes. As the distance between the cells decreases, the adhesion forces increase in magnitude and gradually collision repulsion forces between the cell membranes emerge. These increase in magnitude faster than the \textit{cadherin} adhesion forces (which remain bounded) and an equilibrium between the two types of forces is quickly achieved. The (repulsion) collision forces are not bounded and, if they increase above a particular threshold (corresponding to an extremely small distance between the membranes), the polymerization of the filaments involved in the collision ceases. This ensures that the two cells will not overlap each other. 

The rest of the paper is structured as follows: in Section \ref{sec:fblm} we briefly discuss the FBLM and some of its main components, including the polarization of the lamellipodium and the calibration of the polymerization rate. In Section \ref{sec:adh} we present the new components of the FBLM. We derive in detail the (sub-)model for the collision and adhesion forces, and justify it biologically. In Section \ref{sec:env} we discuss the coupling of the FBLM with the extracellular environment and its response to chemical and haptotaxis stimuli. Finally in Section \ref{sec:exper} we present three numerical experiments. The first two exhibit and compare the effects of cell-cell collision and cell-cell adhesion in the migration and morphology of the cells, and one that exhibits the first stages of cell-cluster formation and its response to a variable chemical and haptotaxis environment. In the final experiment we compare our deduced cell morphologies with the ones of HeLa Cancer cells under \textit{in vitro} cell-cell interaction and migration.

\section{The FBLM.} \label{sec:fblm}

We present here only the main components of the FBLM  and refer to \cite{Oelz2008,Oelz2010a,Schmeiser2010, MOSS-model,MOSS-numeric, Brunk2016, Sfakianakis2018} for more details.

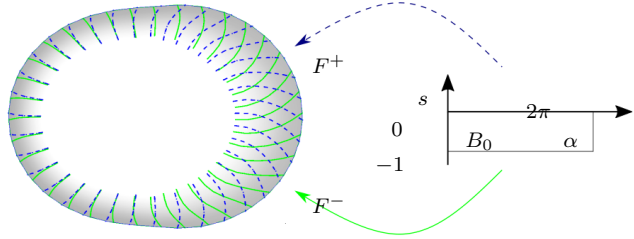
\begin{figure}[t]
	\begin{center}
		\begin{picture}(220,88)
		\put(0,0){\includegraphics[width=22em]{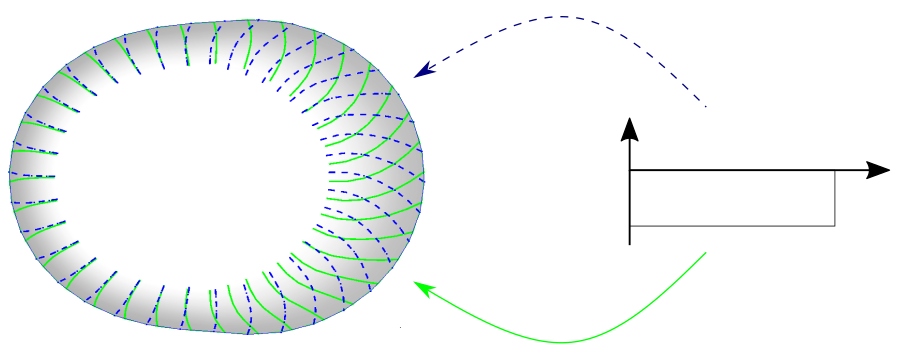}}
		\put(117,65){\scriptsize$F^+$}
		\put(117,12){\scriptsize$F^-$}
		\put(175,37){\scriptsize$B_0$}
		\put(147,41){\scriptsize$0$}
		\put(141,28){\scriptsize$-1$}
		\put(198,48){\scriptsize$2\pi$}
		\put(212,37){\scriptsize$\a$}
		\put(157,53){\scriptsize$s$}
		\end{picture}
	\end{center}
	\caption{Graphical representation of the $\vec F^\pm:B_0\rightarrow \R^2$ mappings that define the lamellipodium.  The $s=0$ boundary of $B_0$ is mapped to the membrane of the cell and the $s=-1$ to the minus-ends of the filaments inside the cell. The filaments and the rest of the functions of $\a$ are periodic with respect to $\a$. The ``filaments'' plotted in the lamellipodium correspond to the discretization interfaces of $B_0$ along the $\a$ direction. The grey color represents the density of F-actin inside the cell.}\label{fig:domains}
\end{figure}

The FBLM is a two-dimensional model that describes the lamellipodium of living cells by including key bio-mechanical processes of the actin-filaments, of the interactions between them, as well as the interactions with the extracellular environment. The basic assumptions behind the FBLM are the following: the lamellipodium is a two dimensional structure, comprised of actin filaments that are organized in two locally parallel families (which are denoted by the superscripts $\pm$). The two families of filaments cover a ring-shaped domain between the membrane of the cell and its interior. In the ``inside'' part of the cell, behind the lamellipodium, further cellular structures are to be found, e.g. \textit{nucleus} and more.  We will henceforth refer to the combined lamellipodium-intracellular space as ``cell'' or ``FBLM-cell'', see e.g. Figure \ref{fig:domains}.

The filaments of the two families are indexed by the continuum variable $\alpha\in [0,2\pi)$, and are parametrised by their arclength 
\begin{equation} \label{eq:arclength}
	\left\{\vec{F}^\pm(\alpha,s,t): -L^\pm(\alpha,t)\le s \le 0\right\}\subset {\mathbb R}^2,
\end{equation}
where  $L^\pm(\alpha,t)$ is the maximal length of the filament $\alpha$ at time $t$. The plus ends of the filaments (at $s=0$) of every family define the outer boundary of the family and ``coincide'' with the membrane of the cell, 
\begin{equation}\label{eq:tether}
\left\{\vec{F}^+(\alpha,0,t): 0\le \alpha < 2\pi\right\} = \left\{\vec{F}^-(\alpha,0,t): 0\le \alpha < 2\pi\right\},  \quad \forall\, t\geq 0 \,.
\end{equation}

For every $(\a,s,t)$ holds that
\begin{equation}\label{eq:inext}
	\left|\partial_s \vec{F}^\pm(\alpha,s,t)\right| = 1 \quad\forall\, (\alpha,s,t) \;.
\end{equation}
This arclength condition can be understood as an \textit{inextensibility} constraint between the  subsequent monomers that comprise the filaments. Moreover, we assume that filaments of the same family do not cross, i.e.
\begin{equation}
	\det \(\partial_\a \vec F^\pm, \partial_s \vec F^\pm\)>0
\end{equation}
and that filaments of different families cross at most once
\begin{equation}
	\Big\{ \forall (\a^+,\a^-)\ \exists \text{ at most one } (s^+,s^-) :\, \vec{F}^+(\alpha^+,s^+,t) = \vec{F}^-(\alpha^-,s^-,t)\Big\}.
\end{equation}

The FBLM is comprised of the force balance system 
\begin{align} 
0=&\underbrace{\mu^B \partial_s^2\left(\eta\, \partial_s^2 \vec{F}\right)}_\subtext{bending} -\underbrace{\partial_s\left(\eta\, \lambda_{\rm inext} \partial_s \vec{F}\right)}_\subtext{in-extensibility}\nonumber\\
&+\underbrace{\mu^A \eta\, D_t \vec{F}}_\subtext{adhesion} \nonumber + \underbrace{\partial_s\left( p(\rho) \partial_\a \vec{F}^{\perp}\right)
	-\partial_\a \left( p(\rho) \partial_s \vec{F}^{\perp}\right)}_\subtext{pressure} \nonumber\\
&\pm\underbrace{\partial_s\left(\eta\,\eta^* \widehat{\mu^T} (\phi-\phi_0)\partial_s \vec{F}^{\perp}\right)}_\subtext{twisting}  
+ \underbrace{\eta\,\eta^* \widehat{\mu^S}\left(D_t \vec{F} - D_t^* \vec{F}^*\right)}_\subtext{stretching} \;,\label{eq:strong}
\end{align}
where $\vec{F}^\bot = (F_1,F_2)^\bot = (-F_2,F_1)$ and where the $\pm$ notation has been dropped here to focus on one of the two filament families. The other family, for which a similar equation holds, is indicated by the superscript $^*$.

The function  $\eta(\alpha,s,t)$ represents the local density of filaments of length at least $-s$ at time $t$ with respect to $\alpha$. Its evolution is dictated, along with $L(\alpha,t)$, by a particular submodel that includes the effects of \textit{actin polymerization}, \textit{filament nucleation}, \textit{branching}, and \textit{capping}. The derivation of this submodel is thoroughly discussed in \cite{MOSS-model}. 

The first term of the FBLM (\ref{eq:strong}) describes the resistance of the filaments against bending, the second term describes the tangential tension force that enforces the inextensibility constraint (\ref{eq:inext}) with the \textit{Lagrange multiplier} $\lambda_{\rm inext}(\alpha,s,t)$,  and the third term describes the friction between the filament and the substrate. The \textit{material derivative} operator
\begin{equation}
	D_t  := \partial_t  - v\partial_s 
\end{equation}
describes the velocity of F-actin relative to the substrate, and $v(\alpha,t)\ge 0$ is the polymerization rate at the leading edge of the filaments. Similarly,  $D_t^\ast  := \partial_t  - v^\ast \partial_s$ is the corresponding material derivative operator for the $^\ast$-family. The pressure term in (\ref{eq:strong}) encodes the Coulomb repulsion between neighbouring filaments of the same family, where the \textit{pressure} $p(\rho)$ is given through the density of actin as
\begin{equation} \label{eq:rho}
	\rho = \frac{\eta}{\left|\det(\partial_\a \vec{F}, \partial_s \vec{F})\right|}  \;.
\end{equation}

The two last terms in (\ref{eq:strong}) model the resistance of the cross-link proteins and branch junctions against changing the inter-filament angle  
\[
	\phi=\arccos (\partial_s \vec{F}\cdot\partial_s \vec{F}^*)
\]
away from the equilibrium angle $\phi_0$, and against stretching.


\begin{subequations}
The system (\ref{eq:strong}) is also subject to the boundary conditions 
\begin{align} 
	- \mu^B\partial_s\left(\eta\partial_s^2 \vec{F}\right) -&~ p(\rho)\partial_\a \vec{F}^\perp + \eta \lambda_{\rm inext} \partial_s \vec{F}  
		\mp\eta\eta^* \widehat{\mu^T}(\phi-\phi_0)\partial_s \vec{F}^\perp \label{eq:newBC}\\
		=&~\left\{
			\begin{array}{l l}
				\eta \left(f_{\rm tan}(\alpha)\partial_s \vec{F} + f_{\rm inn}(\alpha) \vec{V}(\alpha)\right), & \quad \mbox{for 	} s=-L \;,\nonumber\\
				\pm\lambda_{\rm tether} \nu, & \quad  \mbox{for }  s=0 \;,
			\end{array}\right.\nonumber\\
		\eta \partial_s^2 \vec{F} =&~ 0, \qquad \mbox{for } s=-L,0 \;.\label{eq:newBC_2}
\end{align}
\end{subequations}

The right-hand side of \eqref{eq:newBC} describes various forces applied to the filament ends. At $s=0$ (cell membrane), the force in the  direction $\nu$ orthogonal to the leading edge arises from the constraint (\ref{eq:tether}) with the Lagrange parameter $\lambda_{\rm tether}$. The forces at the inner end-point $s=-L$ model the contraction effect of actin-myosin interaction and are directed toward the interior of the cell, refer to \cite{MOSS-model} for details.


\subsection*{Lamellipodium polarization.} 
Fundamental to the motility of the cells is the polarization of the lamellipodium. The effective pulling force becomes stronger in the direction of the wider lamellipodium and the cell migrates accordingly. 

This is also encoded in the FBLM where the maximal filament length $L(\a,t)$ (and hence the local width of the lamellipodium) depends directly on the local polymerization rate $v(\a,t)$. This was previously modelled in \cite{MOSS-model}, where based on the \textit{capping}, \textit{severing}, and \textit{filament nucleation} processes, it was deduced that 
\begin{equation}\label{eq:width_and_v}
	L(\a,t)= -\frac{\kappa_\text{cap}}{\kappa_\text{sev}} + \sqrt{\frac{\kappa_\text{cap}^2}{\kappa_\text{sev}^2} + \frac{2v(\a,t)}{\kappa_\text{sev}}\log{\frac{\eta(0,t)}{\eta_\text{min}}}}.
\end{equation}
Note the monotonic relation between the polymerization rate $v(\a,t)$ and the lamellipodium width $L(\a,t)$. This is used in the FBLM to control the polarization of the lamellipodium and the migration of the cell. 

\subsection*{Adjusting the polymerization rate.}
We account for two different mechanisms that adjust the polymerization rate $v(\a,t)$. The first is the response of the polymerization machinery to \textit{extracellular} chemical signals, as they are perceived by the cell through specialized transmembrane receptors. The second mechanism represents various (unspecified in this work) \textit{intracellular} processes that might cut off, enhance, or otherwise destabilize the polymerization rate, independently of extracellular chemical or other stimuli. 

The first mechanism responds to the density of the chemoattractant $c$ at the plus ends ($s=0$) of the filaments
\begin{subequations}
\begin{equation}\label{eq:chem.pol.a}
	c^\pm (\a,t) = c\( \vec F^\pm (\a,0,t),t\).
\end{equation}
We assume that the polymerization rate  is adjusted between two  biologically relevant minimum and maximum values $v_{\min}$, $v_{\max}$ in the following manner
\begin{equation}\label{eq:chem.pol}
	v^\pm_\text{ext} (\a,t)= v_{\max} - (v_{\max} -v_{\min}) e^{-\lambda_\text{res} c^\pm (\a,t)},
\end{equation}
\end{subequations}
where the coefficient $\lambda_\text{res}$ represents the response of the cell  to changes of the extracellular chemical. The second mechanism describes the response of the polymerization machinery to internal destabilization processes  that might lead to a plethora of phenomena such as persistent or abruptly changing very high or very low polymerization rates,  etc. We understand the biological significance and distinctive functionality of these mechanisms, and employ them both. Overall, the polymerization rate $v^\pm$ is given by
\begin{equation}\label{eq:int.pol}
	v^\pm(\a,t) = \mathcal D_\text{stb}\(v^\pm_\text{ext}(\a,t)\),
\end{equation}
where $\mathcal D_\text{stb}$ describes the internal controlling mechanism that can potentially depend on a large number of cellular processes. 


\section{Cell-cell adhesion and collision.}\label{sec:adh}
\begin{figure}
	\centering
	\includegraphics[width=0.3\linewidth, angle=90]{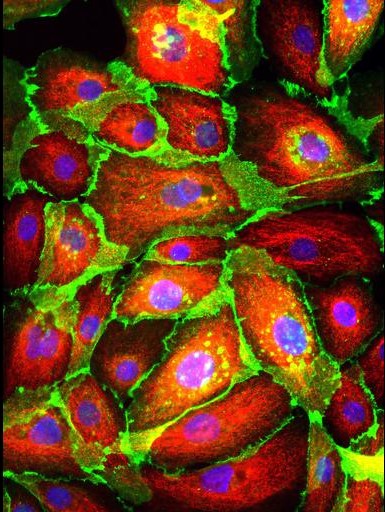}
	\caption{Cryopreserved human mammary epithelial cells stained visualize the calcium-dependent cell-cell adhesion glycoprotein E-cadherin in green. Image by N. Prigozhina (2015) CIL:48102q doi:10.7295/W9CIL48102.}
	\label{fig:mammary}
\end{figure}
The FBLM is developed in a modular way in which every contribution accounts for the potential energy stored in the lamellipodium by the action of the corresponding biological component, see e.g. \cite{Oelz2008,OelzSch-JMB,MOSS-model}. In a similar fashion, cell-cell adhesion and collision are incorporated in the FBLM as additional potential energies acting at the plus-ends of the filaments. To that end we make the following simplifying modelling assumptions:

\textit{Assumption 1:} When two cells come in adhesion proximity (a given parameter of the model), the extracellular domains of their \textit{cadherins} attach and bind with each other. This introduces attractive forces exerted on the plus ends of the  actin-filaments on which the intracellular domain of the \textit{cadherins} are linked to. These adhesion forces increase to a maximum value (a given parameter of the model) with the decrease of the \textit{cadherin} binding length, 

\textit{Assumption 2:} Upon collision, repulsion forces are developed between the two cells and increase rapidly. By nature, these forces can be unbounded, and they soon counteract the effect of the \textit{cadherin} adhesion forces. We model the collision forces pro-actively, i.e. they appear shortly  before the two cells collide (a given distance parameter of the model). Furthermore, to avoid the physical overlapping of the cells, actin polymerization ceases when the collision forces become too large (a given parameter of the model).

We combine the assumptions on adhesion and collision forces, and introduce an attraction-repulsion potential of the form:
\begin{subequations}
\begin{equation}\label{eq:pot.ar}
	U_{ar}[F] = \int_{-\pi}^\pi \eta(\alpha,0,t)\Phi\(\big|F(\alpha,0,t) - \tilde{F}(\alpha,0,t) \big|\) d \alpha,
\end{equation}
where $\tilde{F}(\alpha,0,t)$ is the projection of point $F(\alpha,0,t)$ on the other cells' membrane, and 
\begin{equation}\label{eq:pot.ar.2}
\Phi(r) = \frac{\mu^R}{2}\begin{cases} -(r-r_1)^2+(\frac{1}{r}-r_2)^2, &\quad \text{if } r \leq d_2,\\\
0, &\quad \text{otherwise},
\end{cases}
\end{equation}
where $\mu^R$ represents the intensity of the attraction-repulsion force, $d_2$ is the maximal distance for adhesion attraction and $r_1$ and $r_2$ read:
\begin{equation}\label{eq:pot.ar.3}
\begin{cases}
	r_2 = \frac{1}{1/d_1^2 - 1/d_2^2}(d_1-d_2+\frac{1}{d_1^3}-\frac{1}{d_2^3}),\\
	r_1 = d_1 + \frac{1}{d_1^3}-\frac{r_2}{d_1^2}
\end{cases}.
\end{equation}
\end{subequations}

\begin{figure}
	\center
	\includegraphics[scale=0.32]{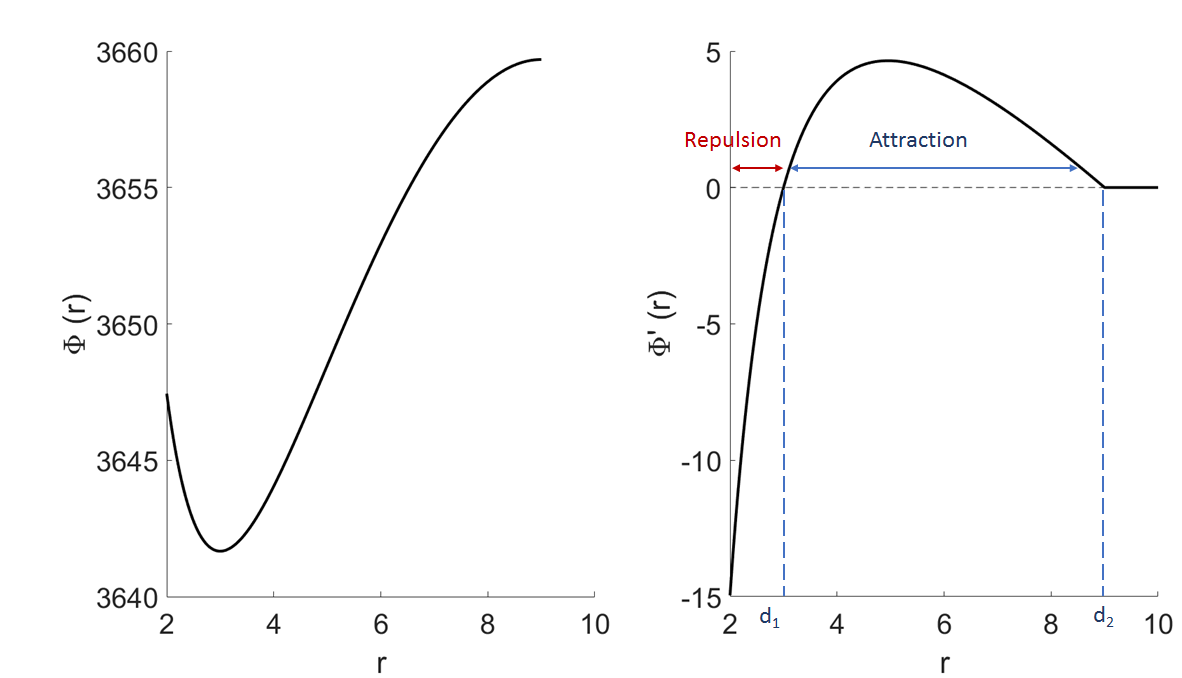}
	\caption{Left: the potential $\Phi(r)$ \eqref{eq:pot.ar.2} for $d_1 = 3$, $d_2 = 9$. Right: the corresponding potential force $\Phi'(r)$ with a cut-off for $r>d_2$.\label{functionPhi}}
\end{figure}

Thus defined, the function $\Phi(r)$ is as depicted in Figure \ref{functionPhi}, $d_1$ being the size of the repulsion zone, $d_2$ the maximal attraction distance. Note that by \eqref{eq:pot.ar} the combined adhesion-collision force is applied on the membranes of the cells and is compactly supported i.e the two cells will only interact as long as their membranes are at distance lower than $d_2$. 

To incorporate this new mechanical feature in the FBLM, we compute the variation of $U_{ar}$ from \eqref{eq:pot.ar}:
\begin{equation}
\delta U_{ar} \delta F = \int_{-\pi}^\pi \eta(\alpha,0,t) \Phi'\(\big|F - \tilde{F} \big|_{(\alpha,0,t)}\)  \frac{\big(F - \tilde{F}\big)(\alpha,0,t)}{|F - \tilde{F} \big|_{(\alpha,0,t)}} \cdot \delta F(\alpha,0,t) d \alpha, 
\end{equation}
and include its contribution in the (membrane) boundary conditions at $s=0$. In effect that Eqs. \eqref{eq:newBC}-\eqref{eq:newBC_2} recast into
\begin{subequations}
\begin{align}
	- \mu^B\partial_s\left(\eta\partial_s^2 \vec{F}\right) - &~p(\rho)\partial_\a \vec{F}^\perp + \eta \lambda_{\rm inext} \partial_s \vec{F}  
		\mp\eta\eta^* \widehat{\mu^T}(\phi-\phi_0)\partial_s \vec{F}^\perp \label{eq:newnewBC}\\
		=&~\left\{
			\begin{array}{l l}
				\eta \left(f_{\rm tan}(\alpha)\partial_s \vec{F} + f_{\rm inn}(\alpha) \vec{V}(\alpha)\right), & \quad \mbox{for 	} s=-L \;,\\
				\pm\lambda_{\rm tether} \nu- \eta \Phi'\(\big|F - \tilde{F} \big|\)  \frac{F - \tilde{F} }{|F - \tilde{F}|}, & \quad  \mbox{for }  s=0 \;,
			\end{array}\right.\nonumber\\
		\eta \partial_s^2 \vec{F} =&~ 0, \qquad \mbox{for } s=-L,0 \;.\label{eq:newnewBC_2}
\end{align}
\end{subequations}

Furthermore, we assume that the polymerization machinery is destabilized by cell-cell interactions. In particular, when the collision repulsion forces become too large (above a given threshold $\Phi^*>0$), we set the local polymerization rate to 0. On the contrary, when the combined adhesion-collision is attractive, we increase the polymerization rate locally. These considerations are supported by biological studies showing the effects of pulling forces on actin polymerization such as in \cite{Kozlov2004}. More specifically, we adjust the polymerization rate locally by setting:
\begin{equation}
	v_*^{\pm}(\alpha) =
		\left\{\begin{aligned}
				0, & \qquad \text{if } \Phi'\(\big|F - \tilde{F} \big|\) \leq -\Phi^*,\\
				3.5 v^{\pm}(\alpha),& \qquad \text{if } \Phi'\(\big|F - \tilde{F} \big|\) \geq 0
		\end{aligned}\right..
\end{equation}
	
\section{Cell-environment interactions.}\label{sec:env}
To account for more biologically realistic situations, we embed the FBLM in a complex and adaptive extracellular environment. The particular coupling of the FBLM with the extracellular environment that we consider here, was previously proposed in \cite{Sfakianakis2018}. We give here a brief description.

We consider an extracellular environment that is comprised of the ECM ---represented by the density of the \textit{glycoprotein} \textit{vitronectin} $v$ onto which the FBLM cells adhere through the binding of the \textit{integrins}---  an extracellular chemical component $c$ that serves as \textit{chemoattractant} for the FBLM cell(-s), and the \textit{matrix degrading metalloproteinases} (MMPs) $m$ that are secreted by the cell and participate in the degradation of the matrix. In our formulation, these environmental components are represented by the \textit{density} of the corresponding (macro-)molecules. Overall the model of the environment reads:
\begin{equation}\label{eq:env}
\left\{\begin{aligned}
	\frac{\partial c}{\partial t}(\vec x,t) &= D_{c} \Delta c(\vec x,t)  + \a\, \mathcal X_{\mathcal P(t)}(\vec x)  - \gamma_1 c(\vec x,t) - \delta_1\, \mathcal X_{\mathcal C(t)}(\vec x)\\
	\frac{\partial m}{\partial t}(\vec x,t) &= D_m \Delta m(\vec x,t) + \beta \mathcal X_{\mathcal C(t)}(\vec x) - \gamma_2 m(\vec x,t)\\
	\frac{\partial v}{\partial t}(\vec x,t) &= -\delta_2 m(\vec x,t) v(\vec x,t)
\end{aligned}\right.
\end{equation}
where $\vec x\in \Omega\subset \mathbb R^2$, $t\geq 0$, $D_c,\, D_m, \a, \beta, \gamma_i, \delta_i \geq 0$, and where $\mathcal X_-$ is the characteristic function of the corresponding set. $\mathcal P$ denotes the support of the pipette(-s) that inject the chemical $c$ in the environment, and the FBLM cell(-s) influence the environment through $\mathcal X_{\mathcal C(t)}(\vec x)$, where $\mathcal C(t)\subset \mathbb R^2$ represents the \textit{full cell} (lamellipodium and internal structures). 


The model of the environment \eqref{eq:env} and the FBLM \eqref{eq:strong} are coupled at three different places: at the characteristic function $\mathcal X_{\mathcal C}$ in \eqref{eq:env}, where the cell $\mathcal C$ produces MMPs and degrades the chemical, at the adhesion coefficient $\mu^A$ in \eqref{eq:strong} which reflects the density of the ECM influences the migration of the cell, and in the polymerization rates $v^\pm_\text{ext}$ of the filaments in  \eqref{eq:chem.pol} which are primarily adjusted according to the density of the extracellular chemical $c$.

Despite the simple structure of the model \eqref{eq:env}, and the numerous biological simplifications we have made, we are able to reconstruct with the FBLM-environment combination, realistic and complex biological phenomena, see e.g. Experiment \ref{exp:cluster}.
	
\section{Experiments and simulations.} \label{sec:exper}
We present three indicative experiments to study the effect of the collision and adhesion components of the FBLM on the migration and morphology of the cells. The first experiment highlights the mechanical effect of  cell-cell collision. In the second experiment, we include the adhesion effect of the \textit{cadherin} protein. In the third experiment we embed several FBLM cells in the same environment and study the first stages of a cell cluster development. In this experiment we also compare our results with a particular biological setting  involving the migration of HeLa cells.

\sectionexp{\textbf{(Cell-cell collision).}}\label{exp:cc.coll}
	We embed two FBLM cells in an environment that it is adhesion and chemically uniform and fixed. Initially, both cells are rotationally symmetric, with diameter 50, and lamellipodia of thickness 8. They are centred at (50,4) and (-50,-4) respectively and the length of their filaments is 10. The environment is such that the adhesion coefficient $\mu^A$ of the FBLM (common for both cells) is uniform and fixed
	\[ \mu^A=0.4101, \]
	and the polymerization rates of the filaments are given by \eqref{eq:chem.pol} and vary in a smooth sinusoidal manner between a minimum $v_\text{min}=1.5$ and a maximum $v_\text{max}=8$ value from the posterior to the anterior side of the cell. The direction of the cell centred at $(50,4)$ is directed eastwards, and of the cell centred at $(-50,-4)$ is  directed westwards. This brings the two cells in a collision path. 

	\begin{figure}
	\centering
	\footnotesize
	\begin{tabular}{ccc}
		\includegraphics[height=9.5em]{{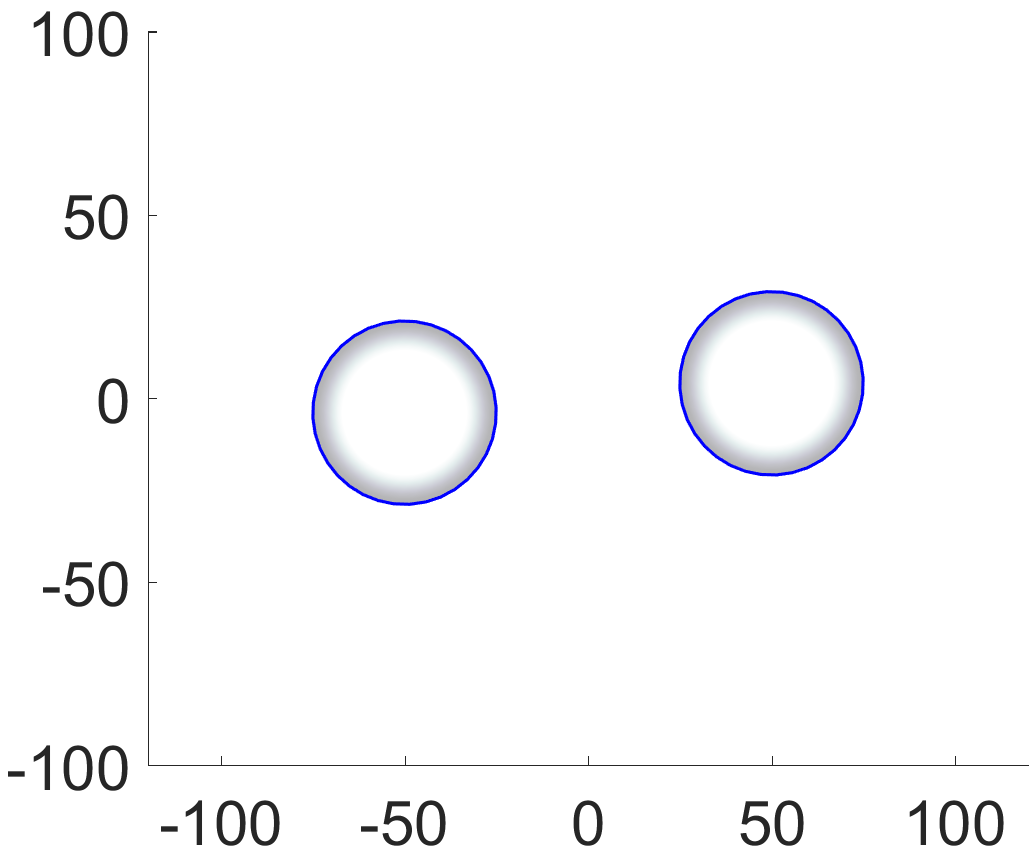}}
		&\includegraphics[height=9.5em]{{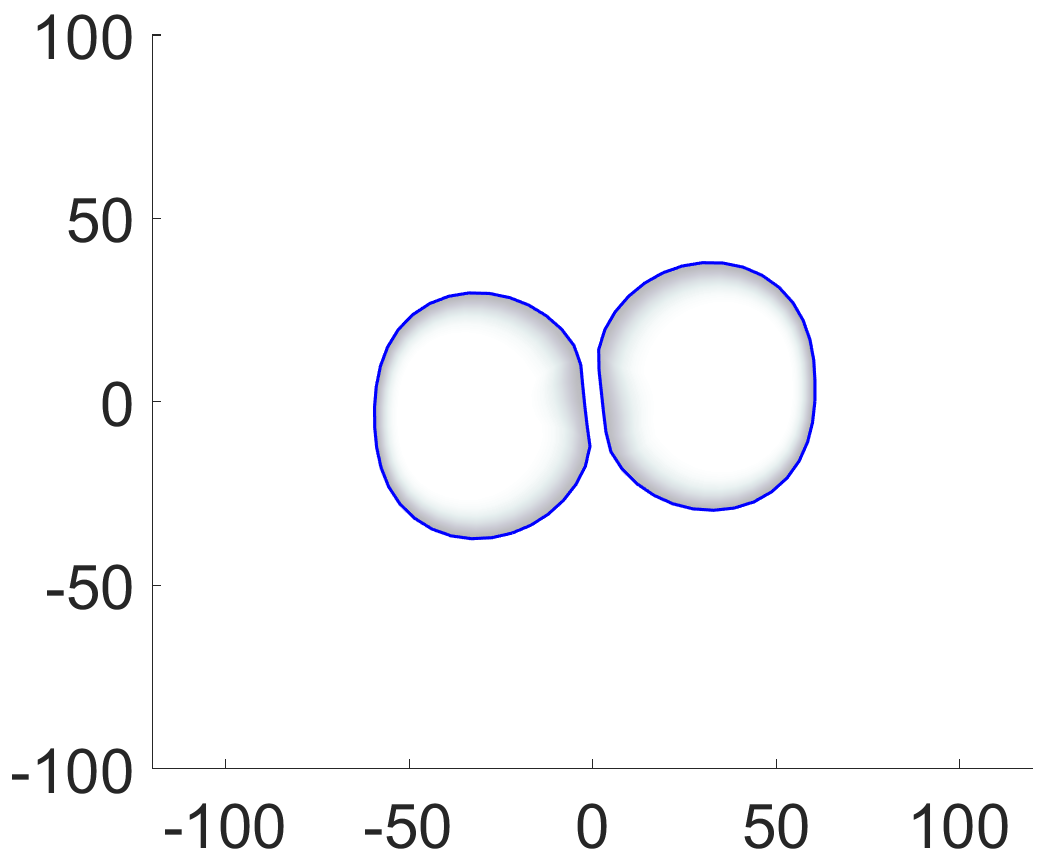}}
		&\includegraphics[height=9.5em]{{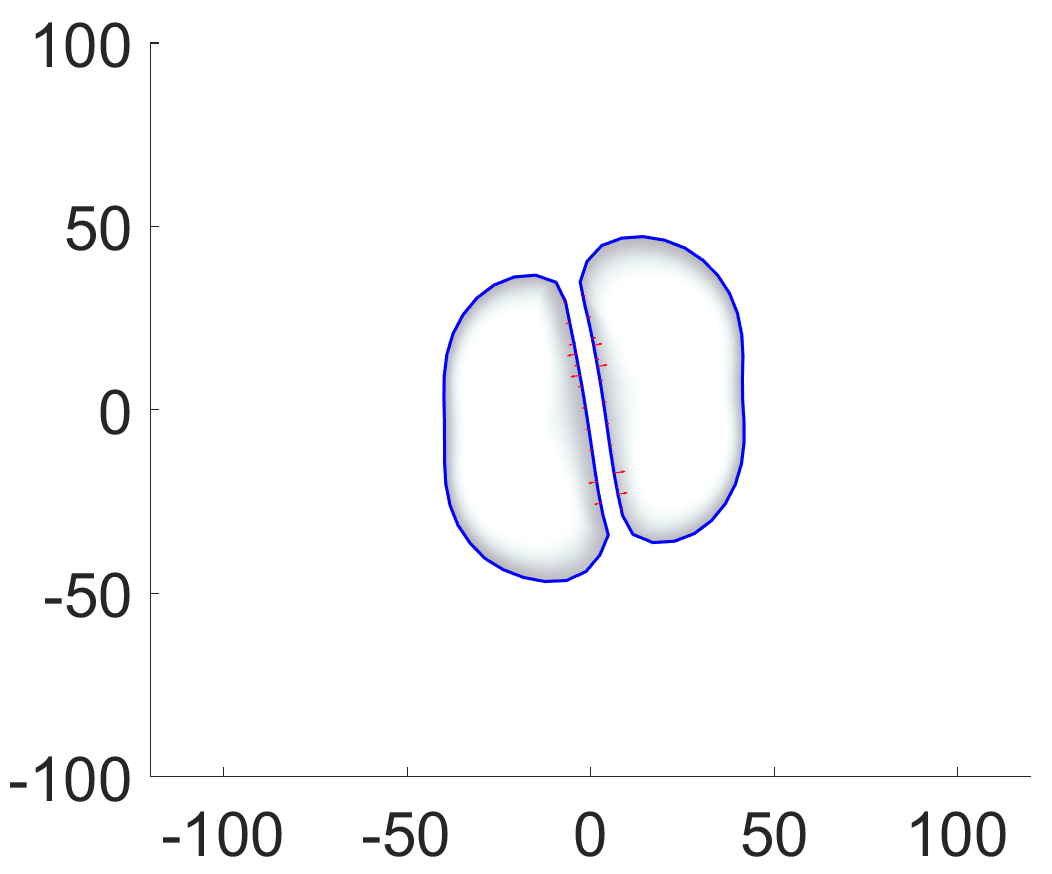}}
		\\
		(i) $t=0.001$ & (ii) $t=5.001$& (iii) $t=10.001$
		\\
		\includegraphics[height=9.5em]{{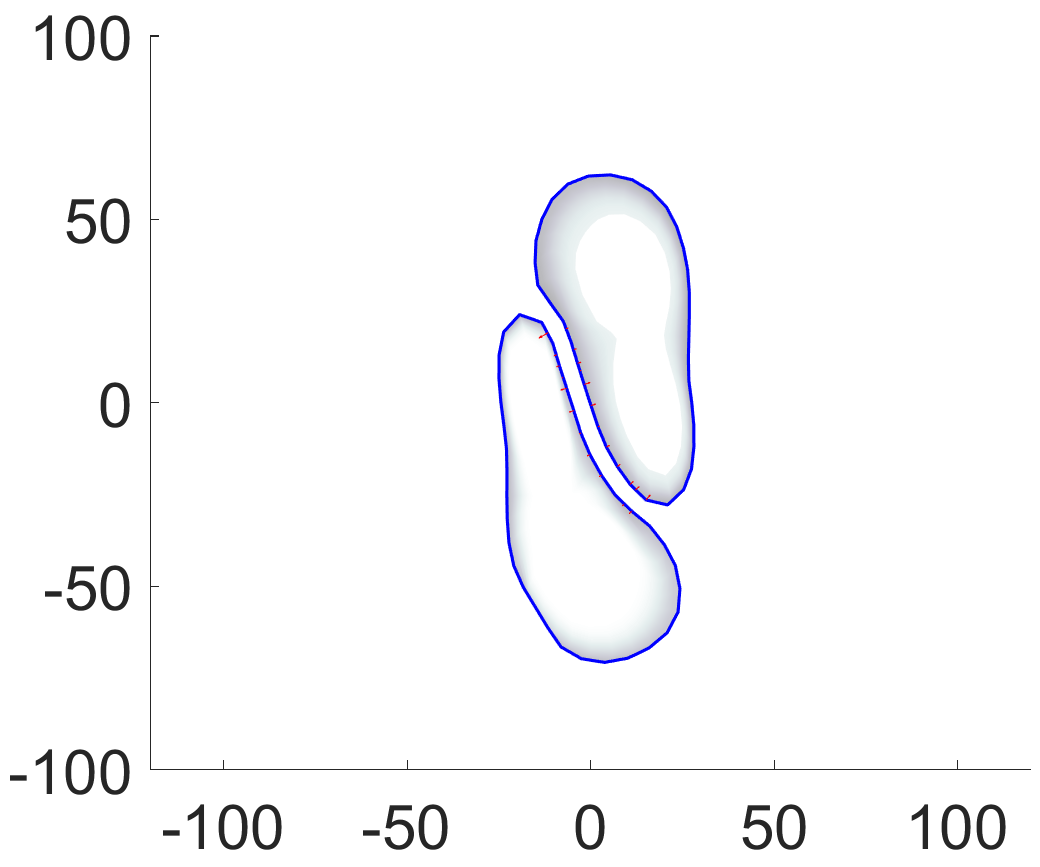}}
		&\includegraphics[height=9.5em]{{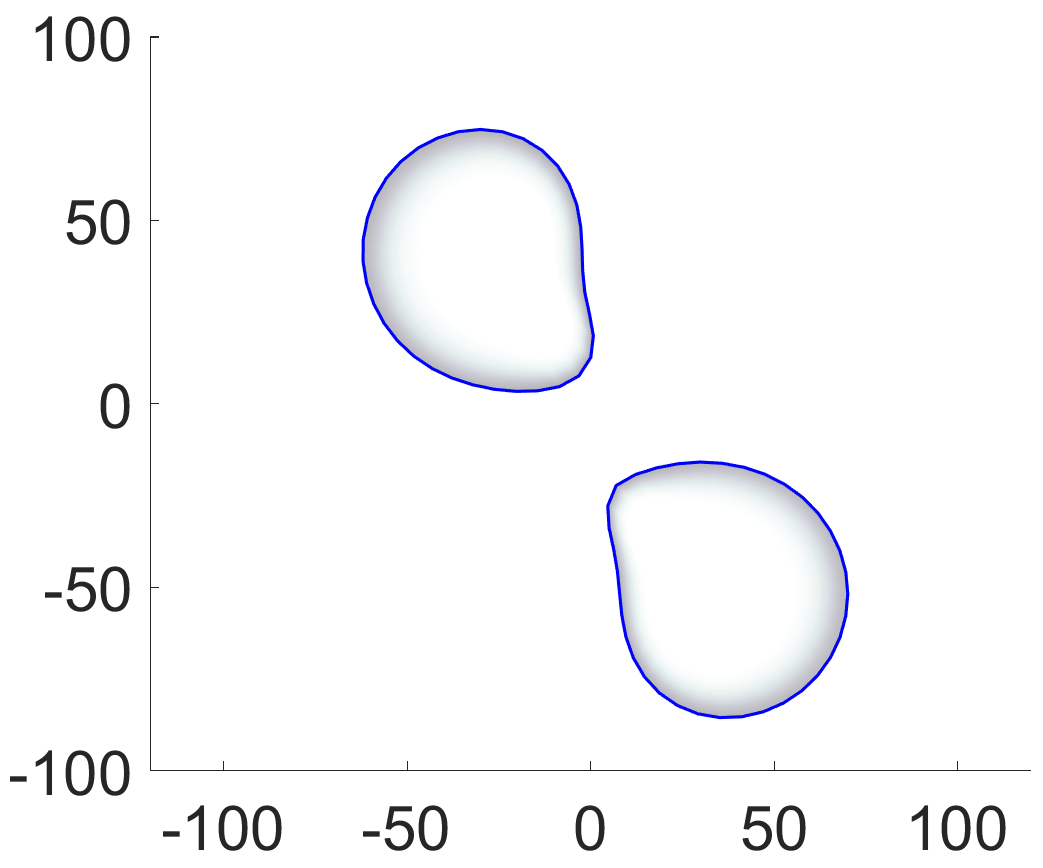}}
		&\includegraphics[height=9.5em]{{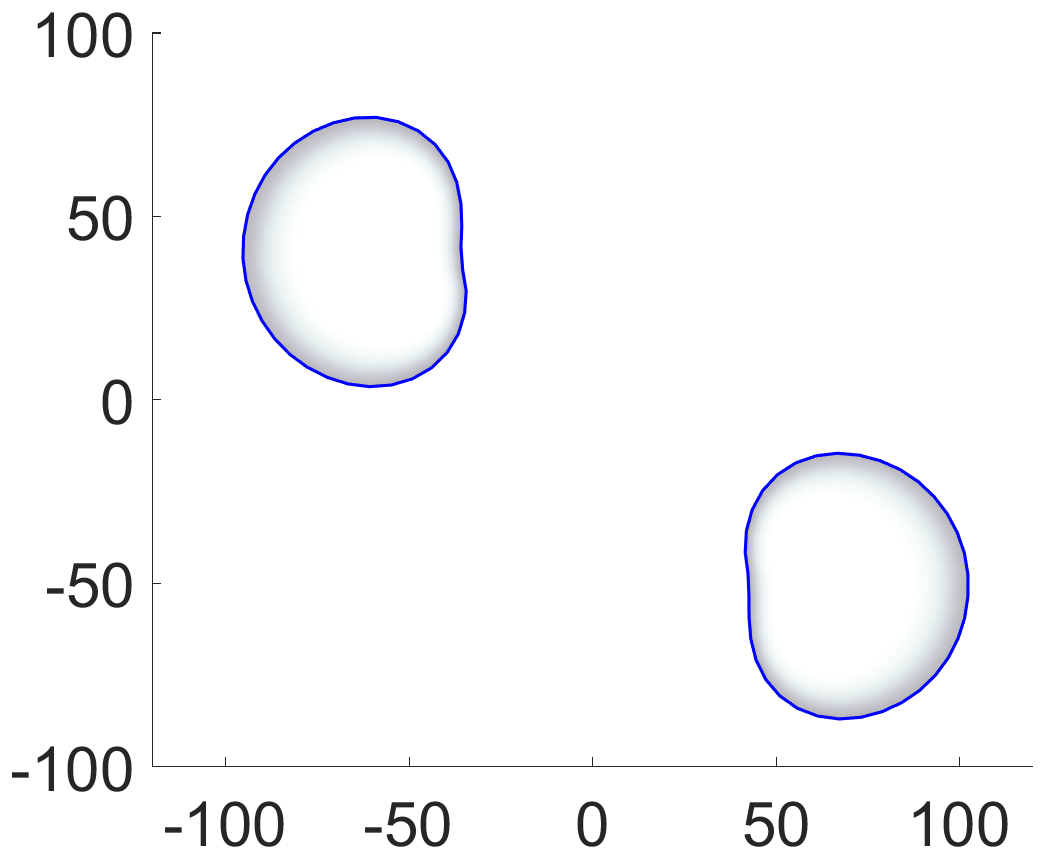}}
		\\
		(iv) $t=20.001$ & (v) $t=30.001$& (vi) $t=37.001$
		\\
	\end{tabular}
	\caption{Experiment \ref{exp:cc.coll} (Cell-cell collision). (i): Two FBLM cells migrate in opposing  east-west directions. (ii)-(iv): The cells collide and deform due to the exchange of repulsive collision forces. The cells slip by each other. (v)-(vi): The deformation of the cells is elastic and the cells recover their pre-collision morphology.}\label{fig:cc.coll}
	\end{figure}
	
	To avoid physical overlapping of the cells, the collision forces act proactively, i.e. when the cells come closer than a pre-defined threshold distance. In this experiment, this distance is set to $5$. When this occurs, the collision forces increase rapidly in magnitude, and when they become very strong (stronger the a predefined threshold), the polymerization of the corresponding filaments ceases. This threshold is set to be $0.01$ in this experiment; the rest of the parameters are given in Table \ref{tbl:fblm.params}.
	
	\begin{SCfigure}[3]
		\centering\footnotesize
		\includegraphics[height=11.5em]{{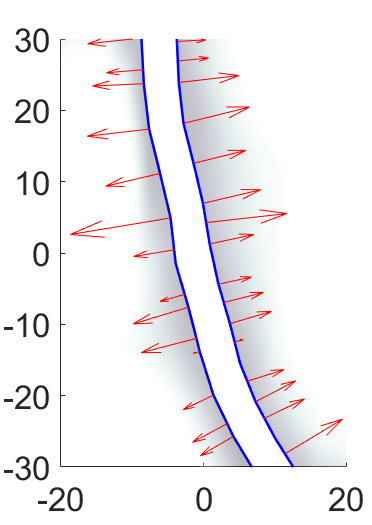}}
		\caption{Experiment \ref{exp:cc.coll} (Cell-cell collision) In a close-up we visualize the repulsive collision forces in action. The magnitude of the forces increases rapidly when the cells come in proximity (closer than a user-defined threshold). When the forces become too large, the polymerizaiton of the corresponding filaments ceases.}\label{fig:cc.coll.detail}
	\end{SCfigure}
	
	In Figure  \ref{fig:cc.coll} we present the corresponding simulation results. After a short time, during which the size of the cells is adjusted to the environmental conditions, the cells collide. The forces that the cells exchange are repulsive and applied symmetrically on the plus ends of the filaments of the two cells; their effect is seen in the deformation of the cells.  When they become very strong, the polymerization of the corresponding filaments ceases. At the non-colliding regions, the polymerization continuous and the cells slip by each other. After moving away from each other, the cells recover the morphology they had before the collision. This implies that the deformation due to collision is \textit{elastic}. This remark can serve as a starting point to  measure the \textit{elastic modulus} of the lamellipodium when cell-type specific experimental evidence are considered.
	
	In Figure \ref{fig:cc.coll.detail} we visualize the force exchange between the two cells. When the distance of the two cells becomes shorter than the (predefined) threshold, the repulsive forces are applied at the plus ends of the corresponding filaments. The magnitude of the forces increases as the distance between the filaments decreases. When the forces reach a maximum value, the corresponding polymerization rates cease. The overall effect is that the cells have the tendency to maintain the threshold distance between then cells.

\sectionexp{\textbf{(Cell-cell adhesion).}}\label{exp:cc.adh}\normalfont
	In this experiment the setting, initial conditions, and the parameters considered, are the same as in the Experiment \ref{exp:cc.coll}. We augment though, this time, the FBLM with  the effect of cadherin forces. These forces are complementary to the cell-cell collision forces and are incorporated in the FBLM in a similar way, see Section \ref{sec:adh}. 
	
	\begin{figure}
	\centering\footnotesize
		\begin{tabular}{ccc}
			\includegraphics[height=9.5em]{{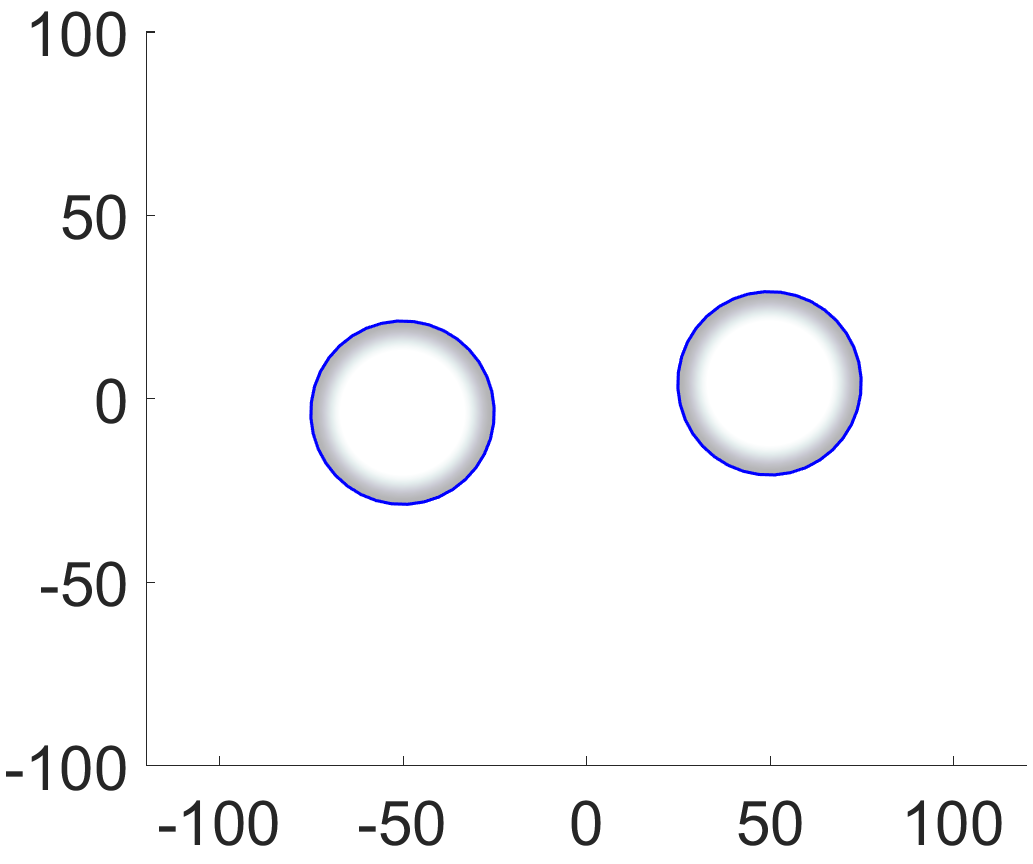}}
			&\includegraphics[height=9.5em]{{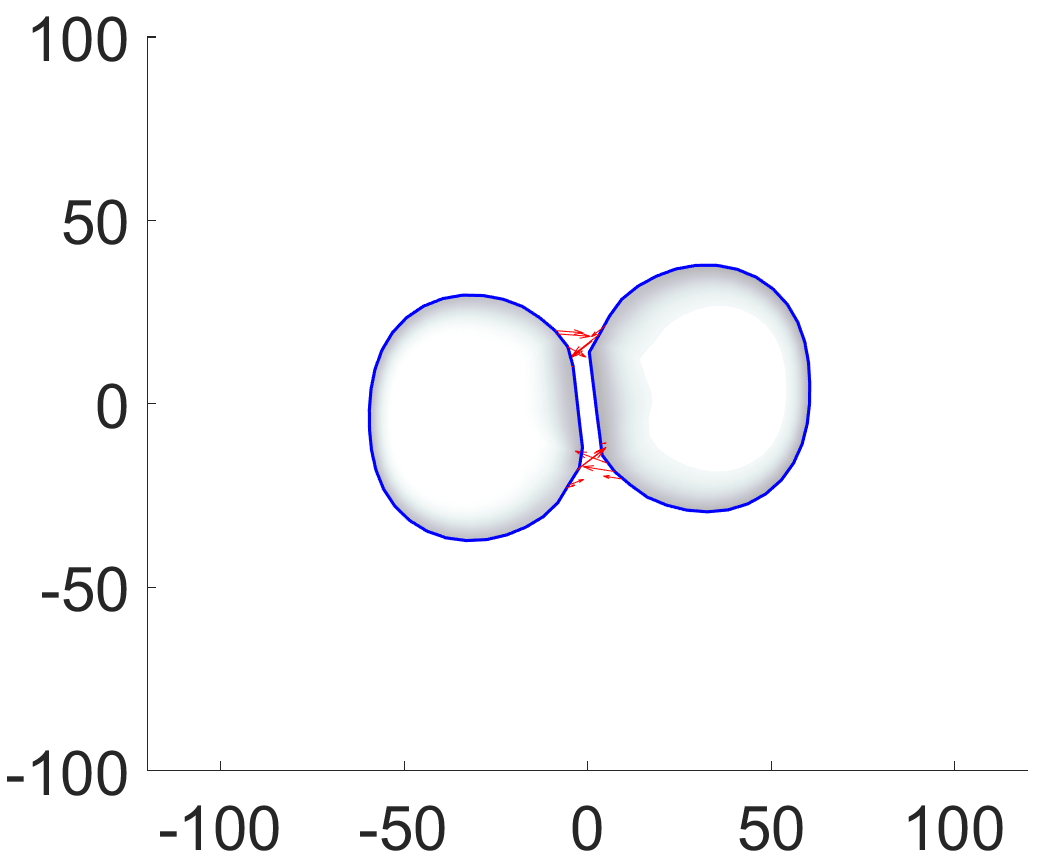}}
			&\includegraphics[height=9.5em]{{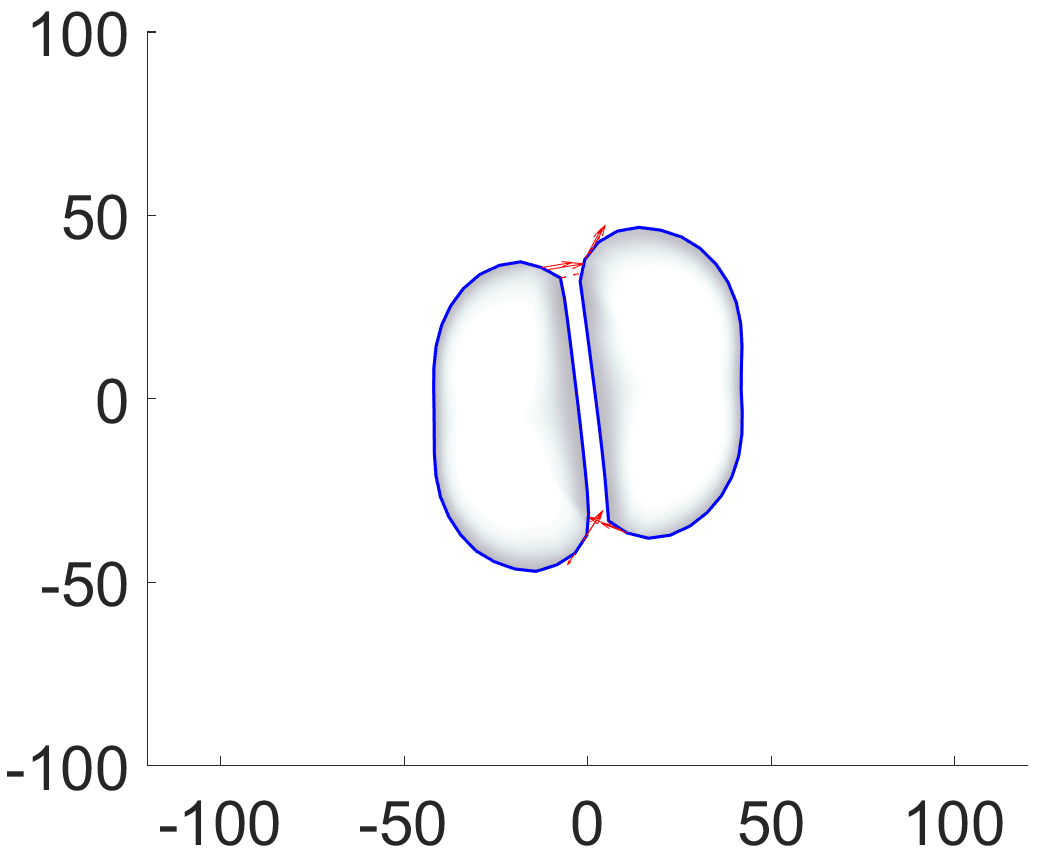}}
			\\
			(i) $t=0.001$ & (ii) $t=5.001$& (iii) $t=10.001$
			\\
			\includegraphics[height=9.5em]{{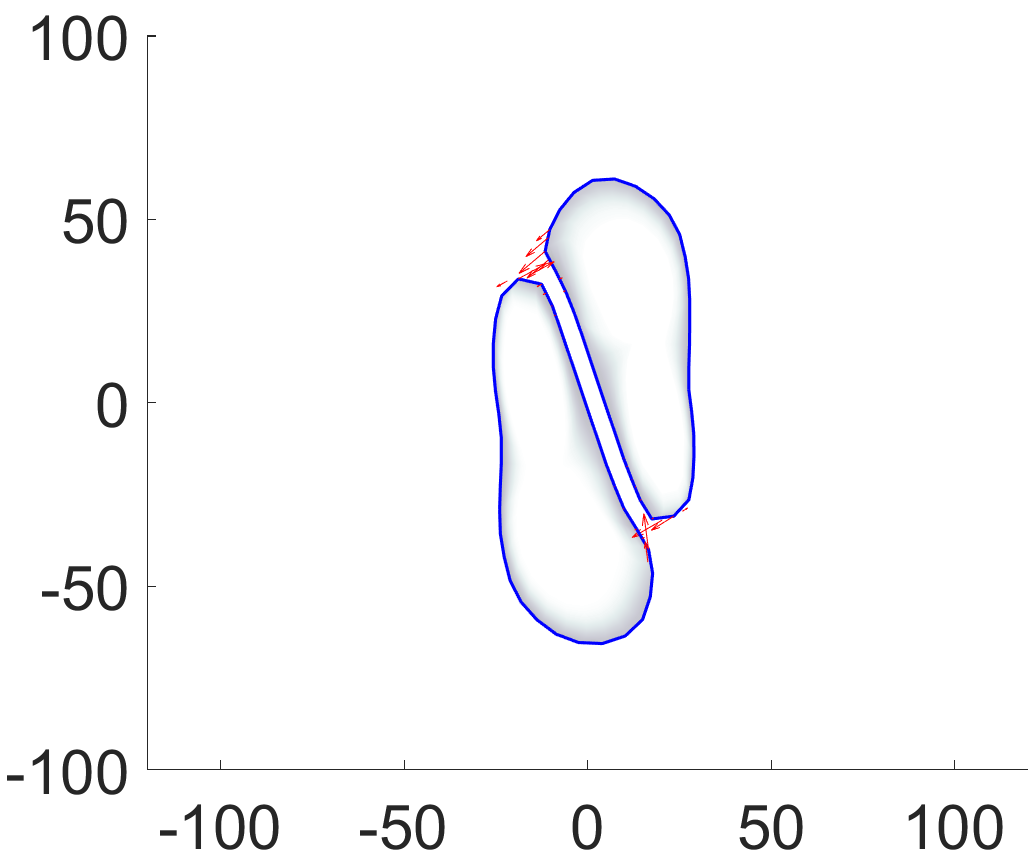}}
			&\includegraphics[height=9.5em]{{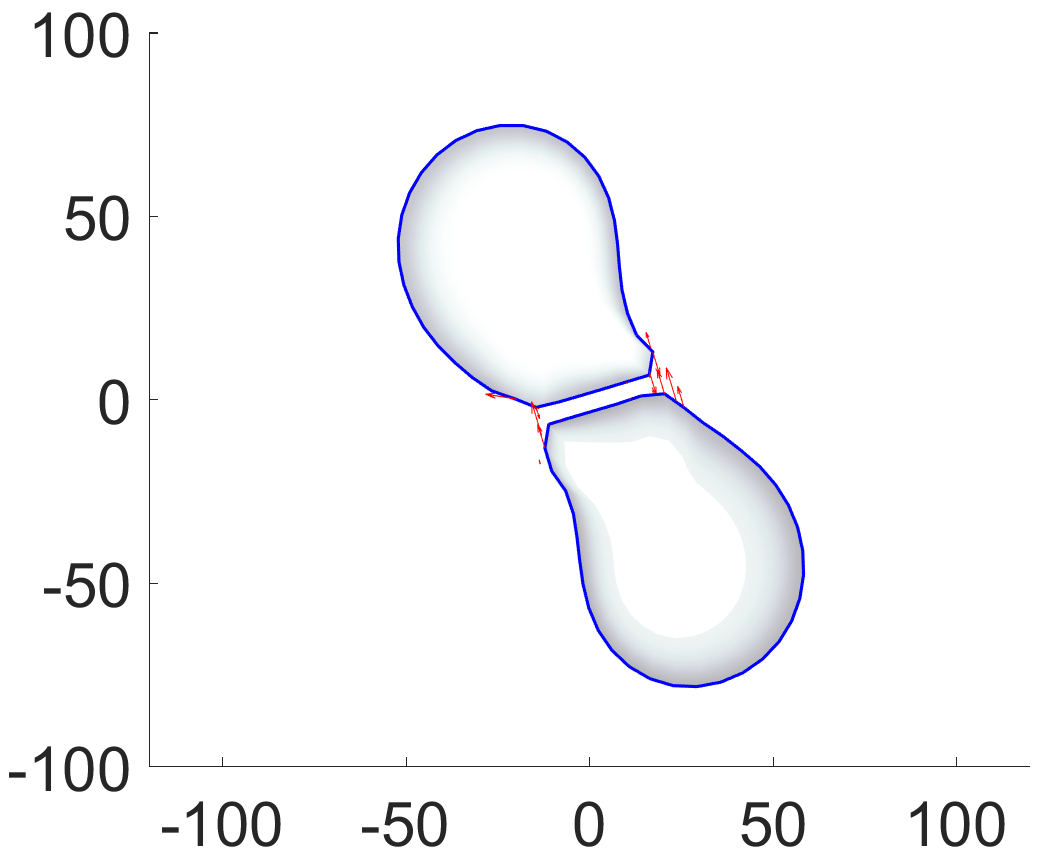}}
			&\includegraphics[height=9.5em]{{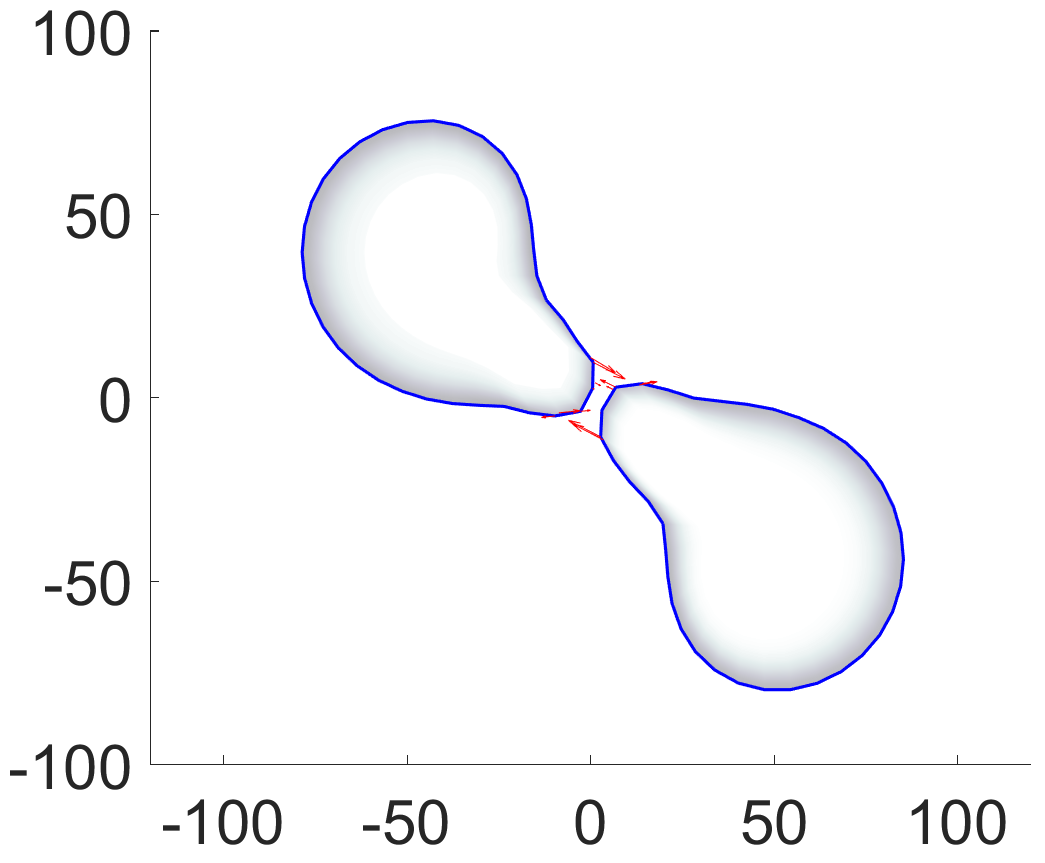}}
			\\
			(iv) $t=20.001$ & (v) $t=30.001$& (vi) $t=37.001$
			\\
			\includegraphics[height=9.5em]{{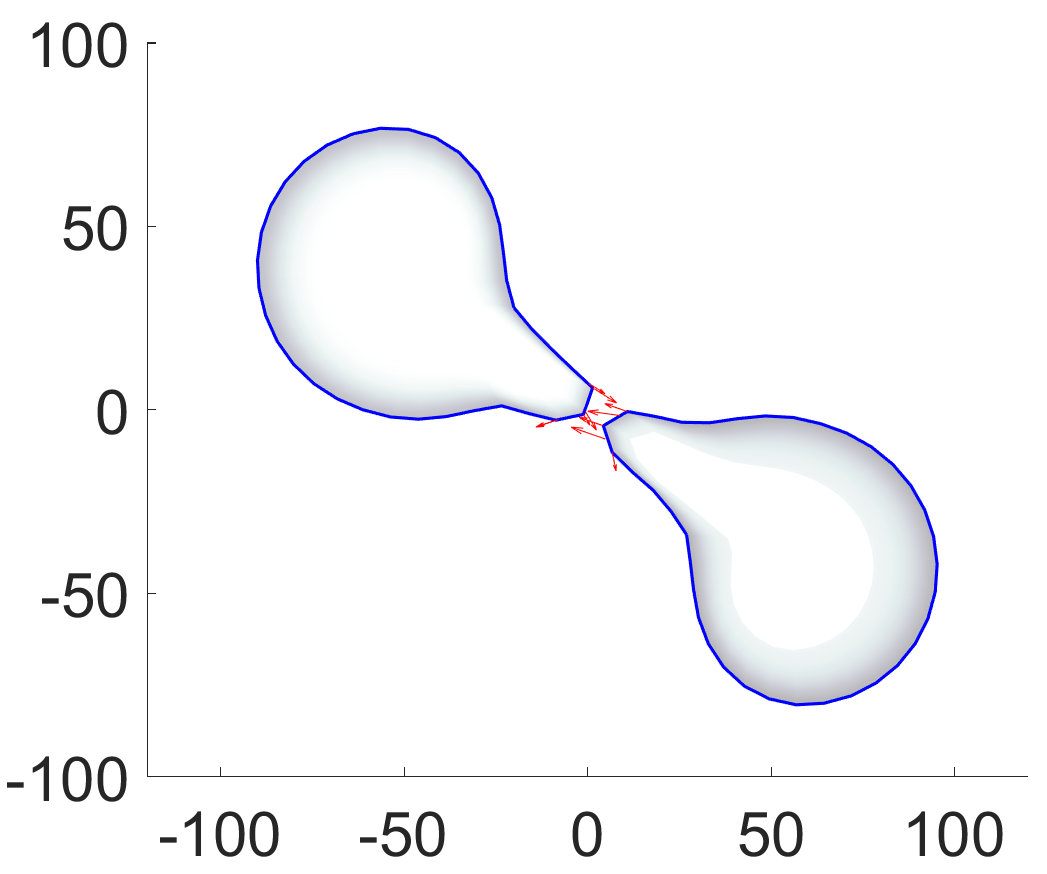}}
			&\includegraphics[height=9.5em]{{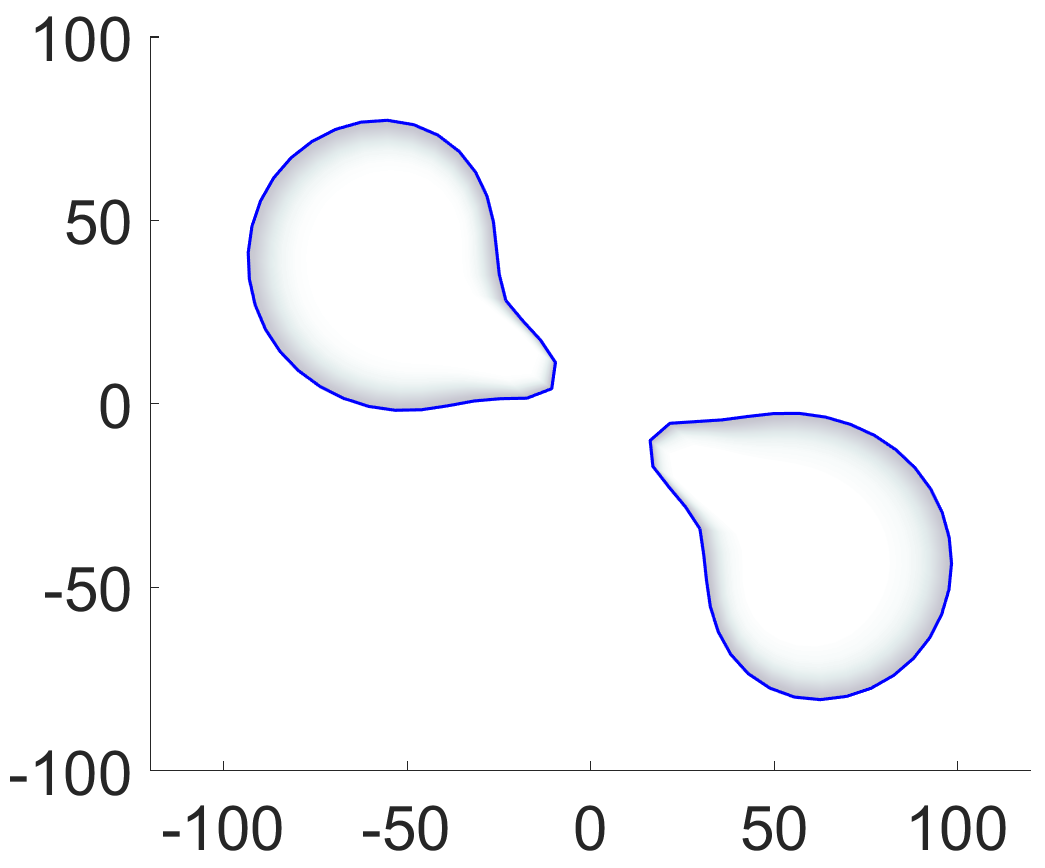}}
			&\includegraphics[height=9.5em]{{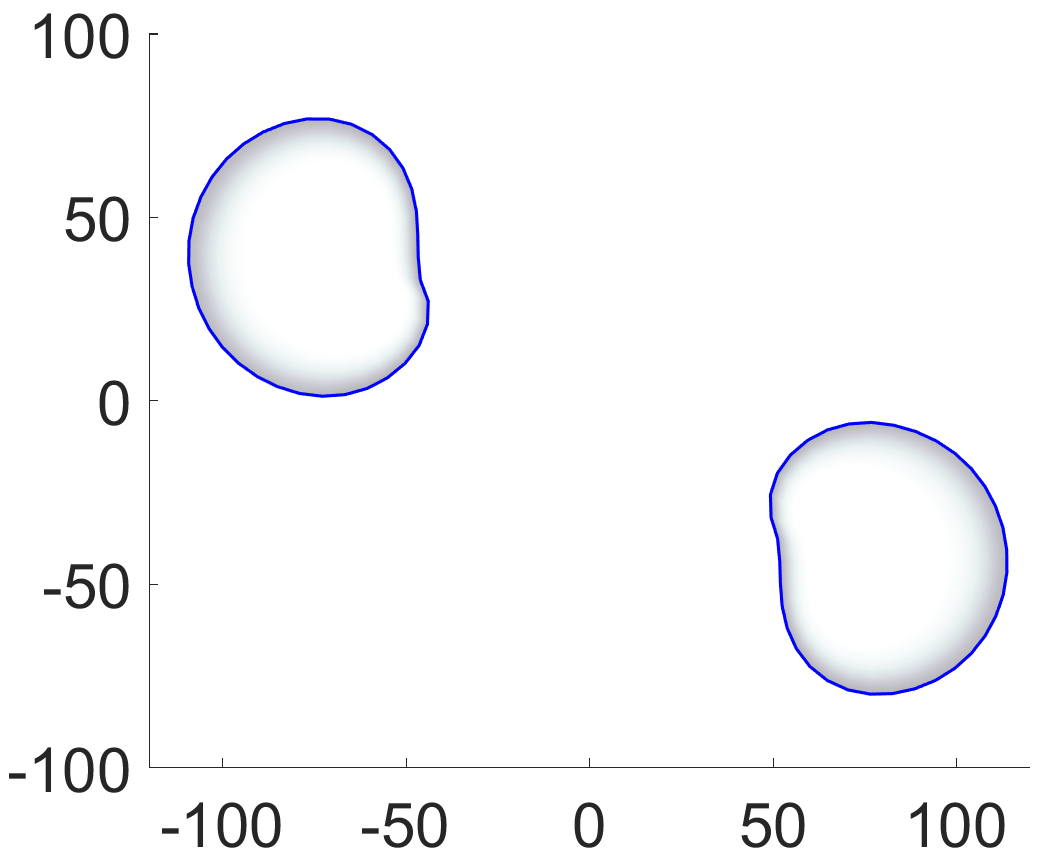}}
			\\
			(vii) $t=40.001$ & (viii) $t=41.001$& (ix) $t=43.001$
		\end{tabular}			
		\caption{Experiment \ref{exp:cc.adh} (Cell-cell adhesion) As in Fig. \ref{fig:cc.coll}, two cells are found in opposing colliding paths. This time though, they are able to develop cadherin induced cell-cell-adhesions. This has an impact in the deformation of the cells, their migrations, their tendency to stick with each other and to resist their separation. (ii)---(v): The adhesive forces are stronger at the ends of the colliding parts of their membranes than the middle parts of it. (vi)---(ix): Note the elastic retraction of the ``tail''/rear part of the cell. (ix): Note also the larger time that is needed for the cells to reach the boundary of the domain, as opposed to the cell-cell collision experiment in Figure \ref{fig:cc.coll}.}\label{fig:cc.adh}
	\end{figure}

	When the distance between the two cells reaches the cell-cell collision threshold, the repulsive collision forces are introduced and counterbalance the attractive adhesion forces. Unless the relative position of the cells change (possibly due to other reasons), the equilibrium between the adhesion and collision forces is maintained. The adhesion threshold distance in this experiment is to $15$, where as the collision threshold distance is set to $5$. When the collision forces become larger than $0.01$ the polymerization of the filaments ceases.
	
	\begin{SCfigure}[2]
		\centering\footnotesize
		\includegraphics[height=11.5em]{{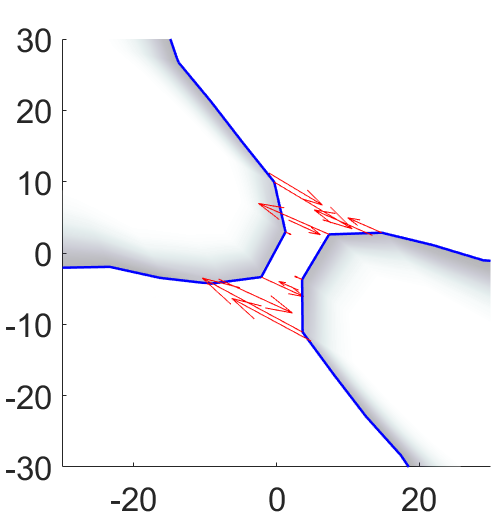}}
		\caption{Experiment \ref{exp:cc.adh} (Cell-cell adhesion). With a close-up in the adhesion zone, we visualize the cadherin adhesion forces. They are exerted at the plus ends of the filaments and are opposite to each other. In the middle region, the adhesion forces have been balanced by the repulsive collision force.}\label{fig:cc.adh.detail}
	\end{SCfigure}
	
	In Figure \ref{fig:cc.adh} we visualize the simulation results of the combined effect of collision and adhesion in the deformation of the cells and their tendency to ``stick together''. It can be seen that at the endo f the contact zones, the adhesion forces are more eminent, whereas in the middle of these zones, no forces are  visible. There, the adhesion and collision forces are in equilibrium. As the cells continue their migration, they slip by each other and their contact zones get stretched due to the adhesion between them. As a result, each cell develops a tail that quickly retracts when the adhesions break.

	We can quantify the effect of cadherin forces, by comparing the average speed of the cells in the two experiment. In the cell-cell collision Experiment \ref{exp:cc.coll}, the cells collide at time $t=5$ at $x=0$ and reach $x=100$ at time $t=37$ i.e. with an average speed $100/(37-5)=3.125$. Similarly, the approximate speed in the cell-cell adhesion case is estimated by $100/(43-5)=2.625$. The difference between the two speeds (although they are not precisely measured) is another effect of the adhesion in the migration of the cells.
	
	In Figure \ref{fig:cc.adh.detail} we visualize a close-up in the tails that the cells develop; there the adhesion forces are clearly visualized. As noted previously, these forces come in pairs, are contractile, and are mostly visible at the ends of the contact zone. The adhesion forces exerted on the filaments in the middle of the zone have been counterbalanced by the repulsive collision forces.


%
%
%

\sectionexp{\textbf{(Cluster formation).}}\label{exp:cluster}
	In the next experiment  we embed several FBLM cells in the same extracellular environment. They collide and adhere with each other, they form a \textit{cell cluster} and we study the first steps of its migration under the influence of an adaptive adhesion and chemical environment.

	\begin{figure}
	\centering
	\footnotesize
	\begin{tabular}{ccc}
		\includegraphics[height=10.2em]{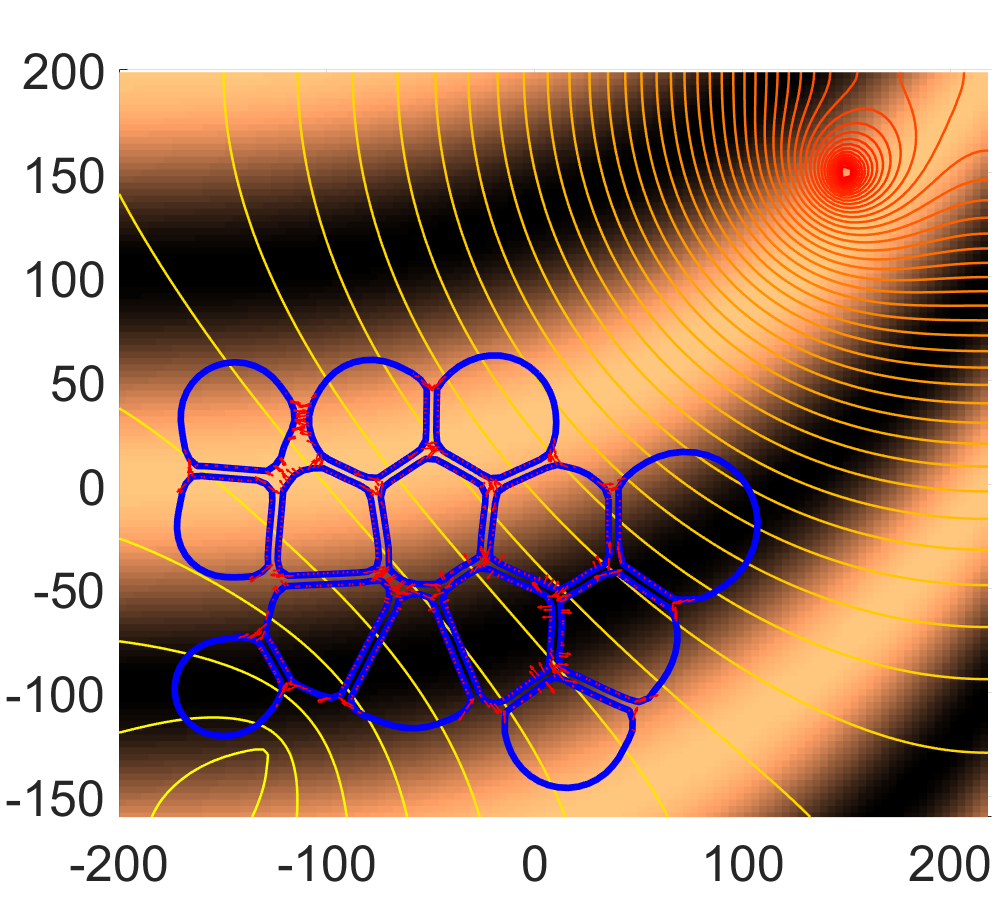}
		&\includegraphics[height=10.2em]{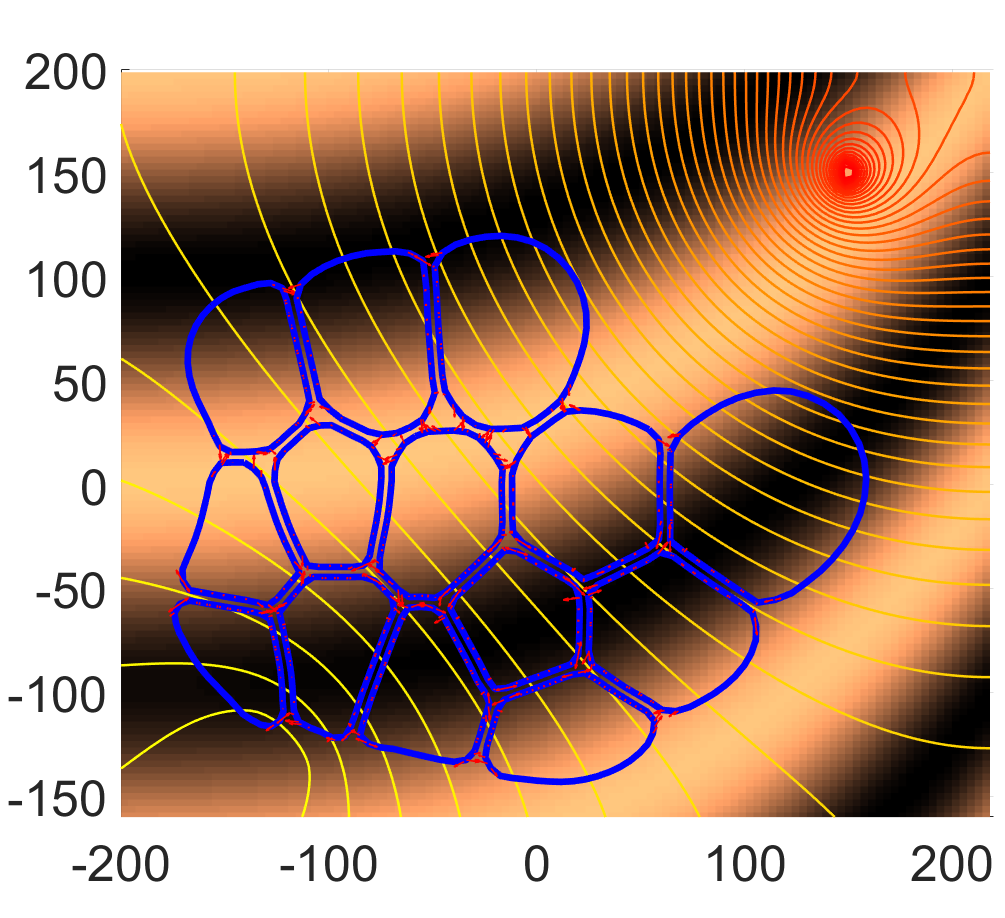}
		&\includegraphics[height=10.2em]{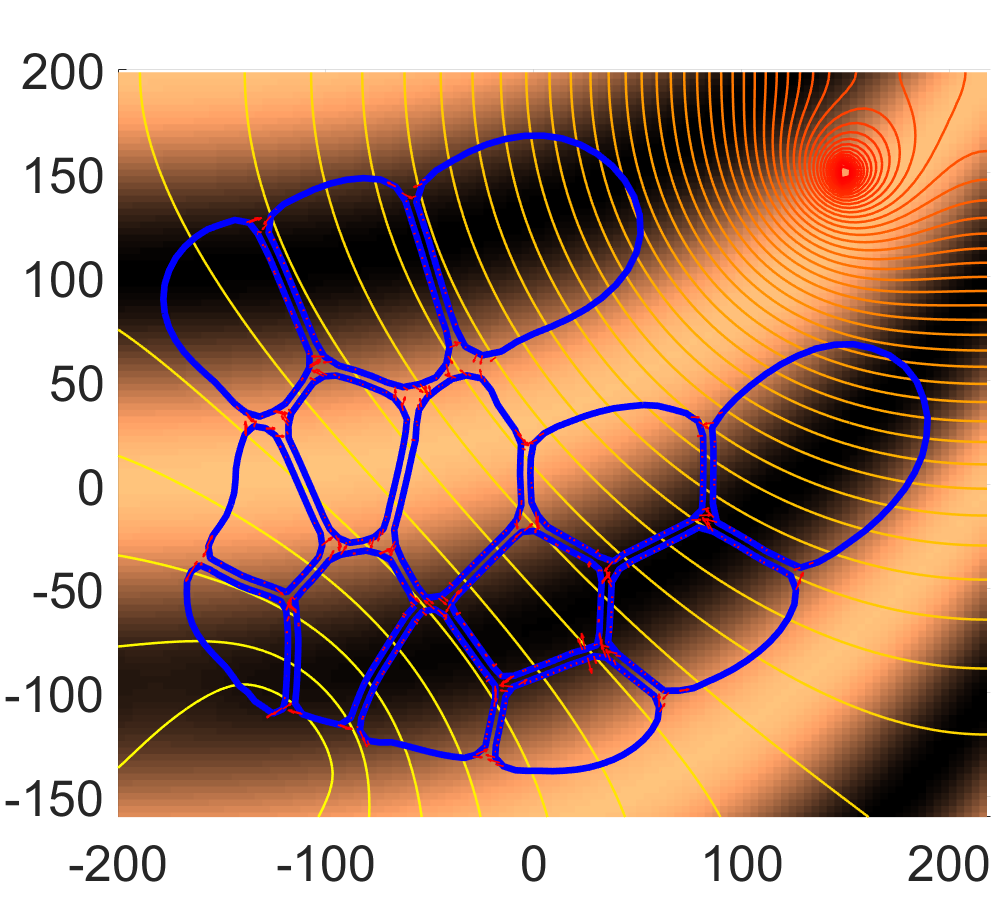}
		~\includegraphics[height=10.2em]{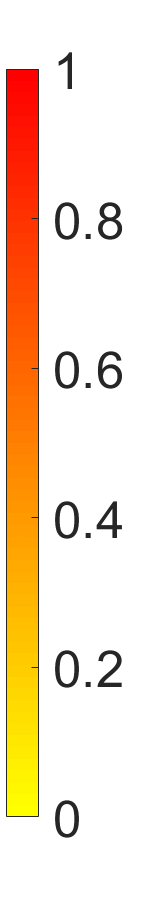}
		\\
		(a) $t=4.308$ & (b) 
		$t=30.004$ & (d) $t=60.006$
		\\
		\includegraphics[height=10.2em]{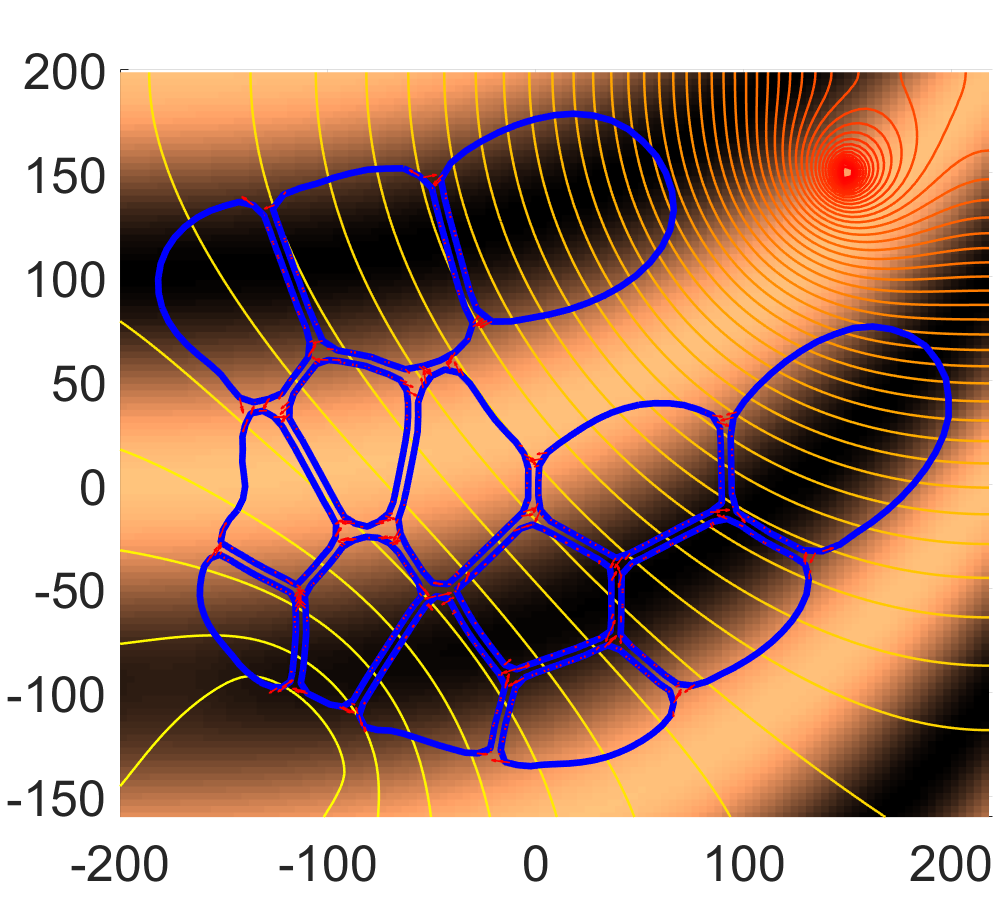}
		&\includegraphics[height=10.2em]{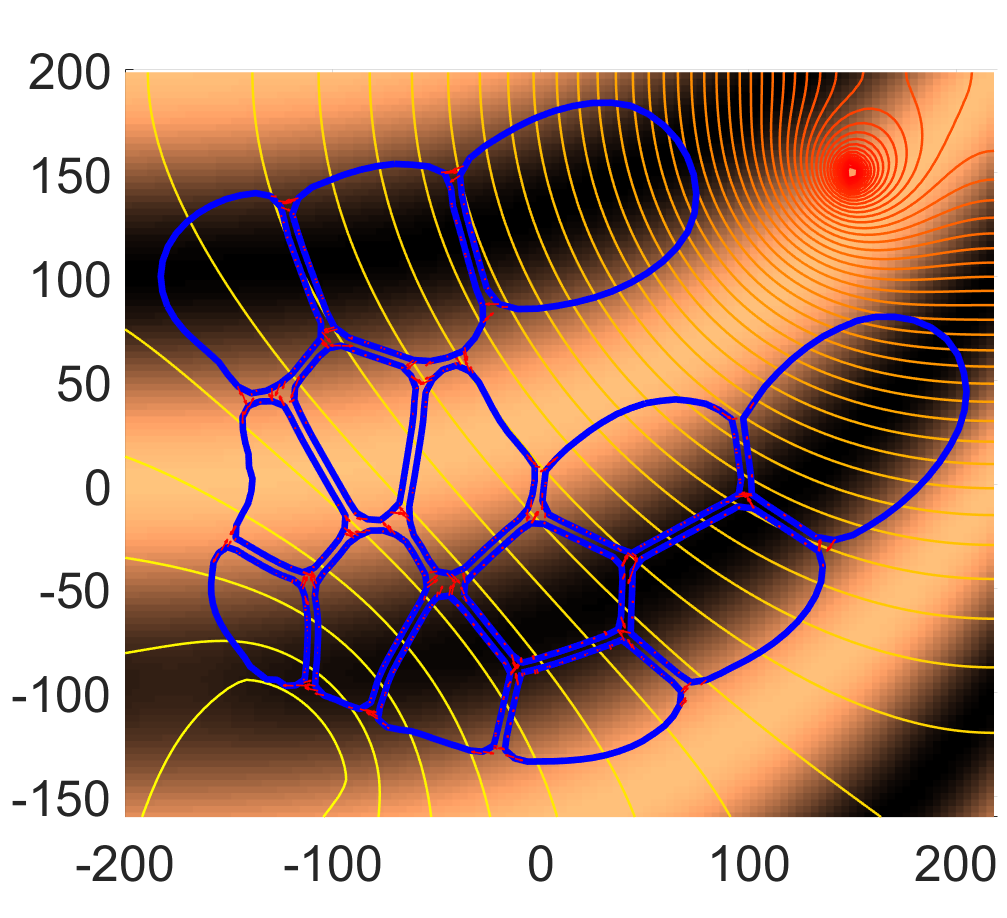}
		&\includegraphics[height=10.2em]{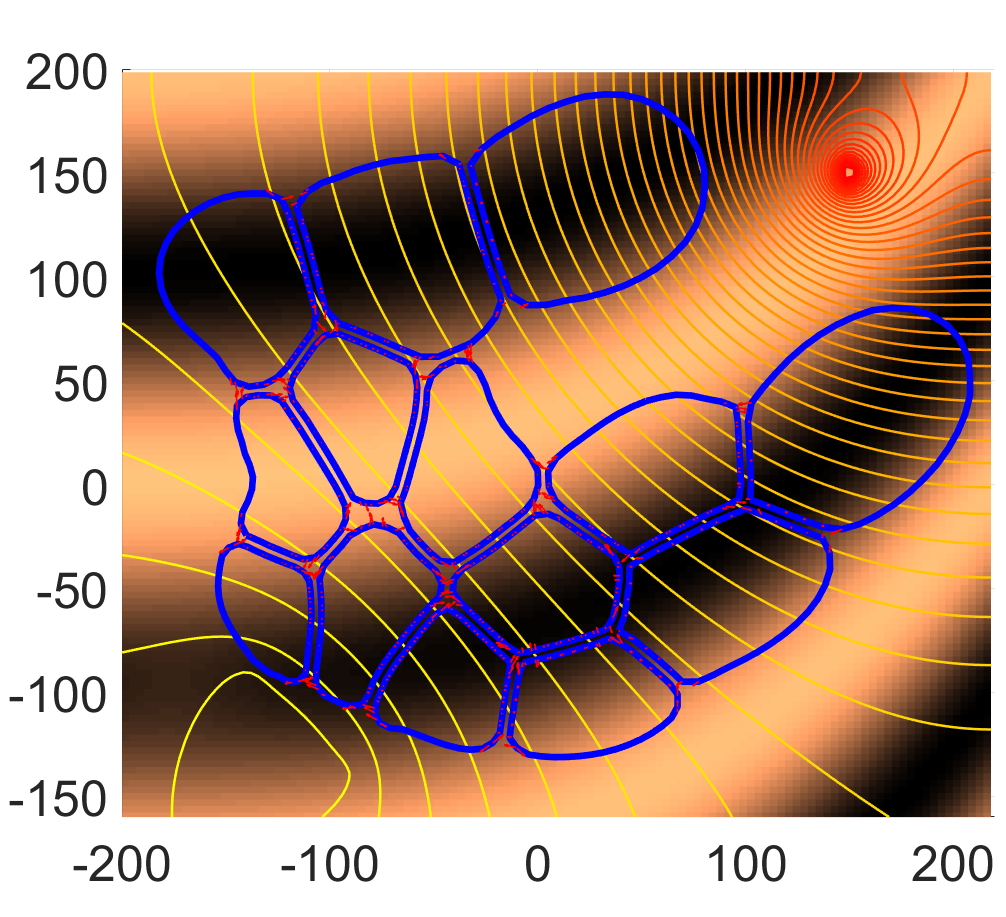}
		~\includegraphics[height=10.2em]{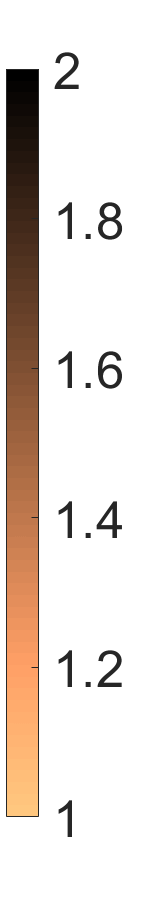}
		\\
		(e) $t=72.517$& (f) $t=81.195$ & (g) $t=90.000$ 
		\\
	\end{tabular}
	\caption{Experiment \ref{exp:cluster} (Cluster formation). A number of 14 FBLM cells are placed in a non-uniform and adaptive environment. The cells collide and adhere with each other, and respond haptotactically to the gradient of the ECM (shown as the background landscape with the corresponding colorbar in the second row) and chemotactically to the chemical gradient (shown as isolines with the colorbar in the first row).}\label{fig:cluster}
	\end{figure}

	We consider 14 cells that are initially the same and rotationally symmetric, and reside in the same extracellular environment. The initial extracellular adhesion landscape and the chemical environment are variable and given respectively by 
	\begin{subequations}
	\begin{align}
			v_0(\vec x)&= \sin^2\(  2\,\frac{x+200}{400}  - 
							 \( \frac{y+150}{350} \)^3 \)\pi+1,\\
			c_0(\vec x)&= e^{-5\cdot10^{-4} \(  10^{-2}(x-30)^2 + (y-40)^2  \) },
	\end{align}
	\end{subequations}
	where $\vec x =(x,y)\in [-200,200]\times[-150,200]$.
	
		\begin{figure}
		\centering
		\footnotesize
		\begin{tabular}{p{0.45\linewidth}p{0.45\linewidth}}
			{\centering \includegraphics[height=11em]{{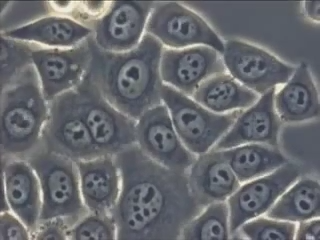}} 
				
			}
			& {\centering \includegraphics[height=11em]{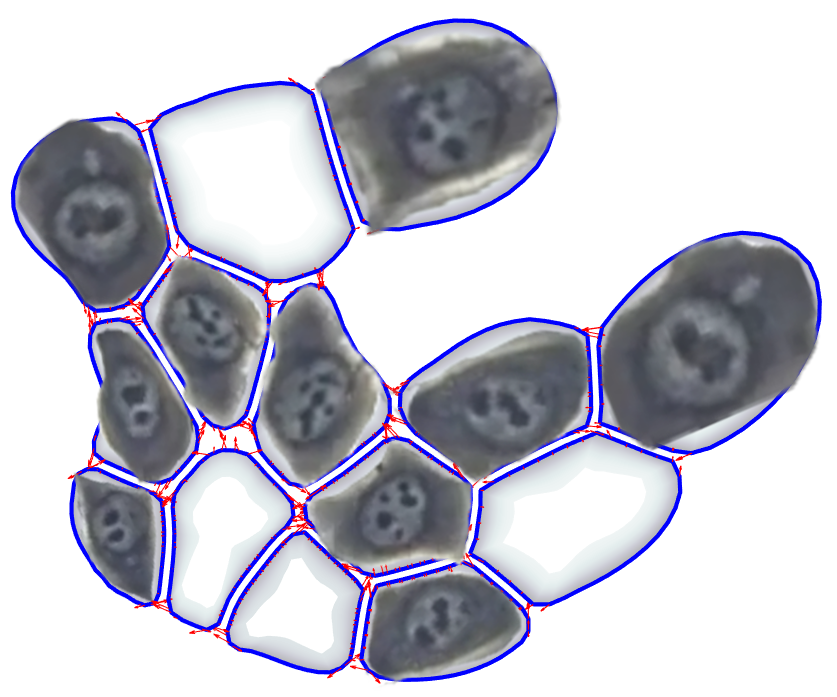} 
				
			}
			\\
			(a) A video frame from \cite{Tlsty2008} shows a number of \textit{in vitro} migrating HeLa cells. We ``extract'' the cells by cutting along their common interfaces. 
			& (b) We superimpose the cut HeLa cells extracted from (a) on the simulation results from Fig. \ref{fig:cluster} (g).
		\end{tabular}
		\caption{Experiment \ref{exp:cluster} (Cluster formation).  We compare the simulation results of Experiment \ref{exp:cluster}, and in particular the morphology of the resulting cells, with \textit{in vitro} culture of the HeLa cancer cells studied in \cite{Tlsty2008}. (a): The single frame from the video in \cite{Tlsty2008} from which HeLa cells were ``extracted''. (b): The fit between the HeLa cells from (a) and our numerical simulations from Fig. \ref{fig:cluster} (g). The comparison follows after properly rotating and scaling the HeLa cells and superimposing them on the simulation results.}\label{fig:cluster.comp}
	\end{figure}
	
	We assume that the cells respond to the chemical and haptotaxis gradients of the environment while at the same time colliding and adhering to each other. The overall model is comprised by 14 FBLM equations of the form \eqref{eq:strong}, one for each cell, and one system for the environment \eqref{eq:env} in which the characteristic function $\mathcal X_{\mathcal C(t)}$, in the degradation of the chemical and the production of the MMPs, is replaced by 
	\[
		\mathcal X_{\cup_i\mathcal C_i(t) },
	\] 
	where $C_i(t)$, $i=1\dots14$ represent the support of the cells, i.e. the area occupied by the  lamellipodium and the inner part of the cells. We assume that all the cells are of the same type and satisfy the FBLM \eqref{eq:strong} with the same parameters; these are given in the Table \ref{tbl:fblm.params}. Their adhesion and collision threshold distances have been set to $15$, $5$, respectively and the collision force threshold to $0.01$. The parameters for the environment \eqref{eq:env} are given in Table \ref{tbl:env.params}.
	
	In Figure \ref{fig:cluster} we present several snapshots of the time evolution of the cluster. The cells respond to the gradient of the ECM $v$, they elongate and align themselves with the higher density of the ECM. The effect of cell-cell adhesion is evident primarily in the cells that are found in the ridges of the ECM. As they are pulled by the neighbouring cells that have already climbed on the higher ECM density regions, they get stretched and elongate in a way ``perpendicular'' to the direction of the ECM. At the same time the cells, and primarily the leading ones, are directed towards the source of the chemical; due to the cell-cell adhesion the whole cluster moves slowly in the same direction.
	
	We do not reproduce in this experiment a particular biological experimental setting. Still the resulting cell morphologies are very close to the biological reality. We exhibit this remark in Fig. \ref{fig:cluster.comp} where we compare our simulation results, taken from Fig. \ref{fig:cluster} (g),  with a particular biological experiment of HeLa cells. In particular, from one frame of the video \cite{Tlsty2008} ---where the time evolution of a (relatively large) cluster of HeLa cells is observed in-vitro--- we ``cut out'' some of the HeLa cells and superimpose them on our simulations.


\begin{table}
	\footnotesize
	\centering
	\begin{tabular}{ c p{0.27\linewidth} |p{0.24\linewidth}| p{0.18\linewidth}}
		
		symb.&\hfill~description\hfill~&\hfill~value\hfill~&\hfill~comment\hfill~\\ 
		\hline
		& & & \\
		$\mu^B$&bending elasticity& $\rm 0.07\,pN\,\upmu m^2$ &\cite{Gittes1993}  \\		
		$\mu^A$&adhesion& $\rm 0.4101\,pN\, min\, \upmu m^{-2}$& \cite{Li2003,Oberhauser2002} \& \cite{Oelz2008,Oelz2010a,Schmeiser2010}\\
		$\mu^T$&cross-link twisting& $\rm 7.1\times 10^{-3}\, \upmu m$ &\\
		$\mu^S$&cross-link stretching& $\rm 7.1 \times 10^{-3}\, pN\, min\, \upmu m^{-1} $& \\
		$\phi_0$&crosslinker equil. angle& $70^o$ & \cite{Schmeiser2010}\\
		$\mu^\text{IP}$&actin-myosin strength&$\rm 0.1\, pN\, \upmu m^{-2}$& \\
		$v_\text{min}$&minimal polymerization&$\rm 1.5\,\upmu m\, min^{-1}$&in biological range\\
		$v_\text{max}$&maximal polymerization&$\rm 8\,\upmu m\, min^{-1}$&in biological range\\
		$\mu^P $ &pressure constant & $\rm 0.1\, pN\, \upmu m $& \\
		$A_0$ & equilibrium inner area & $\rm 650\, \upmu m^2$&\cite{Verkhovsky1999,Small1978}\\
		$\lambda_\text{inext}$&inextensibility&$20$&\\
		$\lambda_\text{tether}$&membrane tethering&$1\times 10^{-3}$&\\
	\end{tabular}
	\caption{Basic set of parameter values used in the numerical simulations of the FBLM in all the experiment of this work. These parameters have been adopted from \cite{MOSS-numeric, Sfakianakis2018}.}
	\label{tbl:fblm.params}      
\end{table}

\begin{table}
	\footnotesize
	\centering
	\begin{tabular}{c p{0.32\linewidth} |p{0.24\linewidth}}
		symb.&\hfill~description\hfill~ &\hfill~value\hfill~\\ 
		\hline
		& &\\[-0.7em]
		$D_c$  &diffusion of the chemical& $3\times10^3\, {\rm cm^2min^{-1}}$\\
		$D_m$ &diffusion of the MMPs& $3\times10^3\, {\rm cm^2min^{-1}}$\\
		$\alpha_1$ &production rate of chemical& $10^2\, {\rm mol\,min^{-1}}$\\ 
		$\beta$ & production of MMPs& $0.1\, {\rm mol\,min^{-1}}$\\
		$\gamma_1$ &decay of the chemical& $10\, {\rm mol\,min^{-1}}$\\
		$\gamma_2$ &decay of the MMPs& $10\, {\rm mol\,min^{-1}}$\\
		$\delta_1$ &degr. chemical by the cell& $10^4\, {\rm mol\,min^{-1}}$\\
		$\delta_2$ &degr. of the ECM by the MMPs&$0\,{\rm cm^2mol^{-1}min^{-1}}$
	\end{tabular}
	
	\caption{Parameter sets used for the simulation of the environment \eqref{eq:env} in the Experiment \ref{exp:cluster} (cluster formation).}
	\label{tbl:env.params}      
\end{table}

\section{Discussion.}
We propose in this work an extension of the actin-based cell motility model \eqref{eq:strong}, termed FBLM, to account also for the collisions and the adhesions between cells. This is achieved by modelling the effect of these two phenomena on the lamellipodium through a single attractive-repulsive potential, \eqref{eq:pot.ar}, which is then incorporated in the  FBLM.

We deduce the adhesion-collision potential \eqref{eq:pot.ar} based on a series of biological assumptions, namely: the adhesion forces are attractive and appear when the cells are in proximity, in a distance justified by the size of the \textit{cadherin} protein. As the distance between the cells decreases, the magnitude of the adhesion forces increases. The adhesion forces can have a maximum value that represents the maximum ``pulling'' strength of the \textit{cadherin} protein. When the cells come in closer proximity, repulsive collision forces appear. The collision forces increase rapidly as the distance between the cells decreases. They are unbounded in magnitude and soon counteract the adhesive effect of the \textit{cadherins}. Both forces are exerted on the plus-end of the filaments and through them are transferred to the cytoskeleton and the rest of the cell. Accordingly, they participate in the $s=0$ boundary conditions of the FBLM, \eqref{eq:newnewBC}.

We study the cell-cell collision and adhesion through three particular experiments: we first simulate the elastic deformation of two cells when only collision is considered. We notice there, the restoration of the cells to their previous morphology after the collision forces cease.  We then incorporate and simulate the effect \textit{cadherins} in the FBLM. We notice the differences in the deformation of the cells as opposed to the collision-only case, the tendency of the cell to ``stick together'' and the elastic retraction fo their ``tails'' when eventually the adhesion forces break. We then embed a number of cells in a non-uniform (haptotaxis and chemotaxis wise) environment while allowing them to collide and adhere with each other. We then compare the results with a \textit{in vitro} experiment of migrating HeLa-cell cluster. We notice the striking similarity of between the simulated and the experimental.

Overall, the cell-cell collision and adhesion extensions of the FBLM that we propose in this paper is of utmost importance for a large number of biologically relevant studies, ranging from \textit{cell-cluster} and \textit{monolayer} formation to \textit{cancer} invasion.

\section*{Acknowledgements.}
	N.S was partially supported by the German Science Foundation (DFG) under the grant SFB 873 ``Maintenance and Differentiation of Stem Cells in Development and Disease". 
	\\The work of C.S. has been supported by the Vienna Science and Technology Fund, Grant no. LS13-029,
	and by the Austrian Science Fund, Grants no. W1245, SFB 65, and W1261.


	\bibliographystyle{plain}
	{	\footnotesize
		\bibliography{main} 
	}


\end{document}